\documentclass[structabstract]{aa}  
\usepackage{graphicx}
\usepackage{txfonts}
\usepackage{natbib}
\begin{document}
   \title{Magnetohydrodynamic  kink waves in two-dimensional 
non-uniform prominence threads}


   \author{I. Arregui
          \inst{1}
          \and
          R. Soler\inst{2}
          \and
          J. L. Ballester
          \inst{1}
          \and
          A. N. Wright
          \inst{3}
          }

   \institute{Departament de F\'{\i}sica, Universitat de les Illes Balears, E-07122, Palma de Mallorca, Spain\\
              \email{[inigo.arregui,joseluis.ballester]@uib.es}
         \and
             Centre for Plasma Astrophysics, Department of Mathematics, Katholieke Universiteit Leuven,
            Celestijnenlaan 200B, 3001 Leuven, Belgium\\
             \email{roberto.soler@wis.kuleuven.be}
             \and
             School of Mathematics and Statistics, University of St. Andrews,  St. Andrews, KY16 9SS, UK\\
              \email{andy@mcs.st-and.ac.uk}
                          }

   \date{Received , ; accepted }

   \abstract
   {}
   {We analyse the oscillatory properties of resonantly damped transverse kink oscillations in two-dimensional prominence threads.}
   {The fine structures are modelled as cylindrically symmetric  magnetic flux tubes with a dense central part with prominence plasma 
properties and an evacuated part, both surrounded by coronal plasma. The equilibrium density is allowed to vary non-uniformly in 
both the transverse and the longitudinal directions. We examine the influence of longitudinal density structuring on periods, damping 
times, and damping rates for transverse kink modes computed by numerically solving the linear resistive magnetohydrodynamic 
(MHD) equations.}
   {The relevant parameters are the length of the thread and the density in the evacuated part of the tube, two 
quantities that are difficult to directly estimate from observations. We find that both of them strongly influence the oscillatory periods 
and damping times, and to a lesser extent the damping ratios. The analysis of the spatial distribution of perturbations and of the 
energy flux into the resonances allows us to explain the obtained damping times.}
   { Implications for prominence seismology, the physics of
resonantly damped kink modes in two-dimensional magnetic flux tubes, and the heating of prominence plasmas are discussed.}


   \maketitle

\section{Introduction}\label{intro}

Quiescent filaments/prominences are cool and dense magnetic and plasma structures suspended against gravity by forces thought 
to be of magnetic origin. In spite of their physical properties, with temperatures and densities that are akin to those in the chromosphere, 
some as yet not well determined mechanisms provide the required thermal isolation from the surrounding coronal plasma and 
mechanical support during typical lifetimes from few days to weeks. The magnetic field that pervades these structures is believed to 
play a key role in the nature and the thermodynamic and mechanical stability of prominences. Early observations carried out with good 
seeing conditions pointed out  that prominences consist of fine threads \citep{dejager59,kuperusTH67}. More recent high-resolution 
H$_\alpha$ observations obtained with the Swedish Solar Telescope (SST) in La Palma \citep{Lin05} and the Dutch Open Telescope 
(DOT) in Tenerife \citep{HA06} have allowed to firmly establish the filament sub-structuring and the basic geometrical and physical 
properties of threads \citep[see also][]{Engvold98, Lin05, Lin08, Lin10}. The sub-structure of quiescent prominences is often composed 
by a myriad of horizontal, dark and fine threads, made of cool absorbing material, believed to outline magnetic flux tubes 
\citep{Engvold98,Engvold08, Linthesis, Lin05, Lin08,Martin08}. The tubes are only partially filled with cool and dense plasma and their 
total length is probably much larger ($\sim$ 10$^5$ km) than the threads themselves. The measured average width of resolved threads 
is about 0.3 arcsec ($\sim$ $210$ km) while their length is between 5 and 40 arcsec ($\sim$ 3500 - 28 000 km).  The absorbing cool 
material is usually visible for up to 20 minutes \citep{Lin05}. The measured widths are close to the current resolution limit, $\sim$ 0.16 arcsec 
at the SST, hence thinner structures are likely to exist.

Small amplitude oscillations in prominence threads are frequently observed \citep[see reviews by][]{OB02,Engvold04,Wiehr04,Ballester06,banerjee07,Engvold08,Oliver09,Ballester10}. Early two-dimensional observations of filaments 
\citep{YiEngvold91,Yi91} revealed that individual threads or groups of them oscillate with  periods that range between 3 and 20 minutes. 
Recent relevant examples are traveling waves propagating along a number of threads with average phase speed of 12 km s$^{-1}$, 
wavelength of 4 arcsec, and oscillatory periods that vary from 3 to 9 minutes \citep{Lin07}, the both propagating and standing oscillations 
detected over large areas of prominences  by \citet{Terradas02} and \citet{Linthesis}, as well as observations from instruments onboard 
space-crafts, such as SoHO \citep{Blanco99,Regnier01,Pouget06} and Hinode \citep{Okamoto07,Terradas08hinode,Ning09}. The transverse 
oscillation nature of some of these events has been clearly established by \citet{Lin09} by combining H$_\alpha$ filtergrams in the plane of the 
sky with H$_\alpha$ Dopplergrams which allow to detect oscillations in the line-of-sight direction. A recurrently observed property of prominence 
oscillations is their rapid temporal damping, with perturbations decaying in time-scales of only a few oscillatory periods \citep{Landman77,Tsubaki86,Tsubaki88,Wiehr89,Molowny-Horas99,Terradas02,Linthesis,Ning09}.

Transverse thread oscillations are commonly interpreted in terms of standing or propagating magnetohydrodynamic (MHD) kink waves. 
The measured periods are of the order of a few minutes and the wavelengths are in between 3000 - 20 000 km, although \citet{Okamoto07} 
report larger wavelengths more consistent with the standing wave interpretation. 
The measured wave quantities  allow us to derive phase speeds that are consistent with the kink speed in magnetic and plasma configurations with
 typical properties of prominence plasmas. The MHD wave interpretation of thread oscillations has allowed the development of theoretical 
 models \citep[see][for recent reviews]{Ballester05,Ballester06}. \citet{JNR97,diaz01,diaz03} considered the MHD eigenmodes supported by a filament thread modelled in Cartesian geometry. More realistic 
studies using cylindrical configurations have extended the initial investigations \citep{diaz02,DR05,DR06}. These studies have determined the 
frequencies and confinement properties of the perturbations as functions of the length and the width of the threads.  Theoretical damping 
mechanisms have also been developed \citep[see][for recent reviews]{Ballester10,AB10}. A systematic  comparative study
of different mechanisms has been presented by \citet{solerthesis}, who assesses the ability of each mechanism to reproduce the observed 
attenuation time-scales. The considered mechanisms include non-ideal effects, such as radiation and thermal conduction, partial ionisation 
through ion-neutral collisions, ion-electron collisions, and  resonant absorption due to coupling to Alfv\'en and slow waves.
Non-ideal effects do not seem to provide the required attenuation time-scales for kink oscillations 
\citep{Ballai03,Carbonell04,Terradas01,Terradas05,Soler07a,Soler08}.  \citet{solerthesis} finds that resonant absorption in the Alfv\'en 
continuum is the only mechanism able to produce the observed attenuation time-scales  
\citep[see][for details]{Soler08,soler09picyl,soler09rapi,soler09slowcont}.  Resonant damping was first considered in this context by 
\citet{Arregui08thread}, and has been subsequently studied in combination with damping in the slow continuum and partial ionisation by \citet{soler09picyl,soler09rapi,soler09slowcont}. These studies have considered one-dimensional models with a density variation in the 
transverse direction only. On the other hand, theoretical studies that take into account the longitudinal density structuring, hence 
incorporating the fact that magnetic tubes supporting threads are only partially filled with cool plasma, consider piece-wise homogeneous models 
in the transverse direction, and hence rule out resonant damping. 

A recent investigation by \citet{soler102dthread} is the exception. These authors have obtained  analytical and semi-analytical approximations 
for periods, damping times, and damping ratios of standing kink modes in a two-dimensional prominence thread model.  However, the 
analysis by \citet{soler102dthread} is restricted to the thin tube and thin boundary approximations, radial inhomogeneity is constrained to the dense cool 
part of the tube only, and the density in the evacuated part of the tube is taken equal to the coronal density, further limiting the prominence threads 
that can be theoretically modelled. These  limitations are removed in the present investigation. We adopt a fully inhomogeneous two-dimensional 
prominence thread model. The density distribution is allowed to vary non-uniformly in both the transverse and longitudinal directions. We also  add another 
relevant parameter with seismological implications, namely, the density in the evacuated part of the tube. By combining the two 
parameters that characterise the longitudinal structuring in prominence threads, the length of the thread and the density in the evacuated part of the 
tube, a wide range of prominence threads with very different physical conditions can be modelled and their oscillatory properties characterised. In 
addition, a general parametric study is numerically performed, that allows us to go beyond the thin tube and thin boundary approximations.

Besides obtaining the influence of longitudinal structuring on periods and damping times for kink modes in two-dimensional threads, and extracting conclusions about their implications for prominence seismology, our study aims at explaining the obtained results.  For this reason, we have performed the analysis of the spatial distribution of perturbations and the energy flux into the resonances. These two analyses, which are novel in the context of prominence oscillations, provide us with a comprehensive explanation for the obtained parametric results and add further insights to the physics of resonantly damped kink modes in two-dimensional equilibrium states.

The paper is organised as follows. Section~\ref{sect2}  describes the thread model, the relevant parameters introduced by its two-dimensional character, the 
linear MHD wave equations to be solved, and the numerical method used for that purpose. Section~\ref{sect3} presents our analysis and results. We first show 
computations of periods, damping times, and damping ratios as a function of the length of the thread and the density in the evacuated part of the tube. Implications for the determination of physical parameters in prominences are discussed. Next,  a qualitative explanation of the obtained results is given by analysing the spatial distribution of perturbations. Finally, we describe our energy analysis, that in combination with the spatial structure of eigenfunctions fully explains the obtained damping times.  In Section~\ref{sect4}, our conclusions are presented.

\begin{figure*}
\begin{center}
 \includegraphics[width=0.8\textwidth]{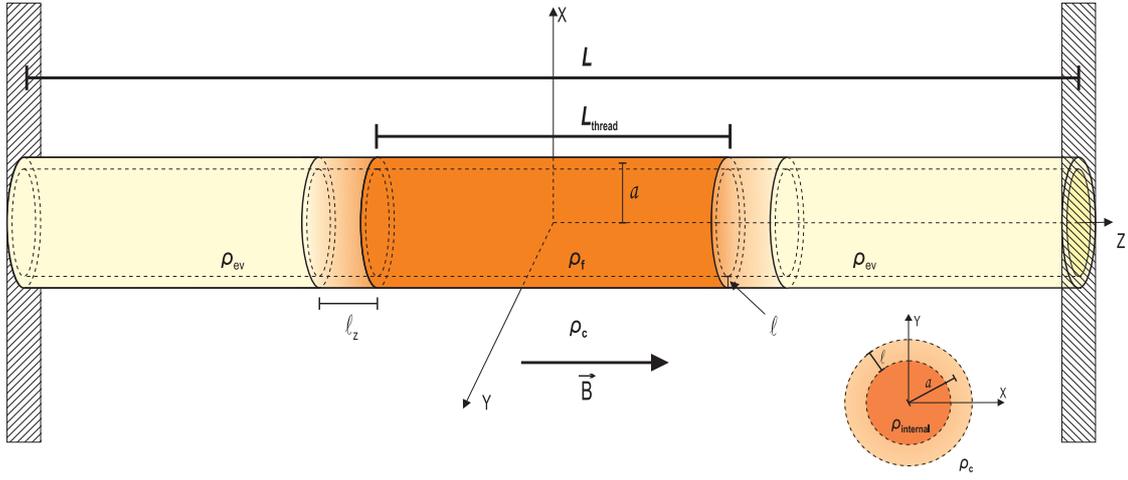}
\end{center}
 \caption{Schematic representation of the model configuration adopted in this work. The two-dimensional cylindrically symmetric line-tied structure consist of a dense part, the thread, with length 
 $L_{\rm thread}$  and density $\rho_{\rm f}$, and two evacuated parts, with density $\rho_{\rm ev}$. Both regions are separated by a non-uniform layer of width $l_{\rm z}$ in the longitudinal direction. Transverse non-uniformity is considered on a layer of width $l$. The structure is surrounded by plasma with coronal properties and  density $\rho_{\rm c}$. \label{modelfig}}
\end{figure*}

\begin{figure*}
 \includegraphics[width=5.8cm,height=4.55cm]{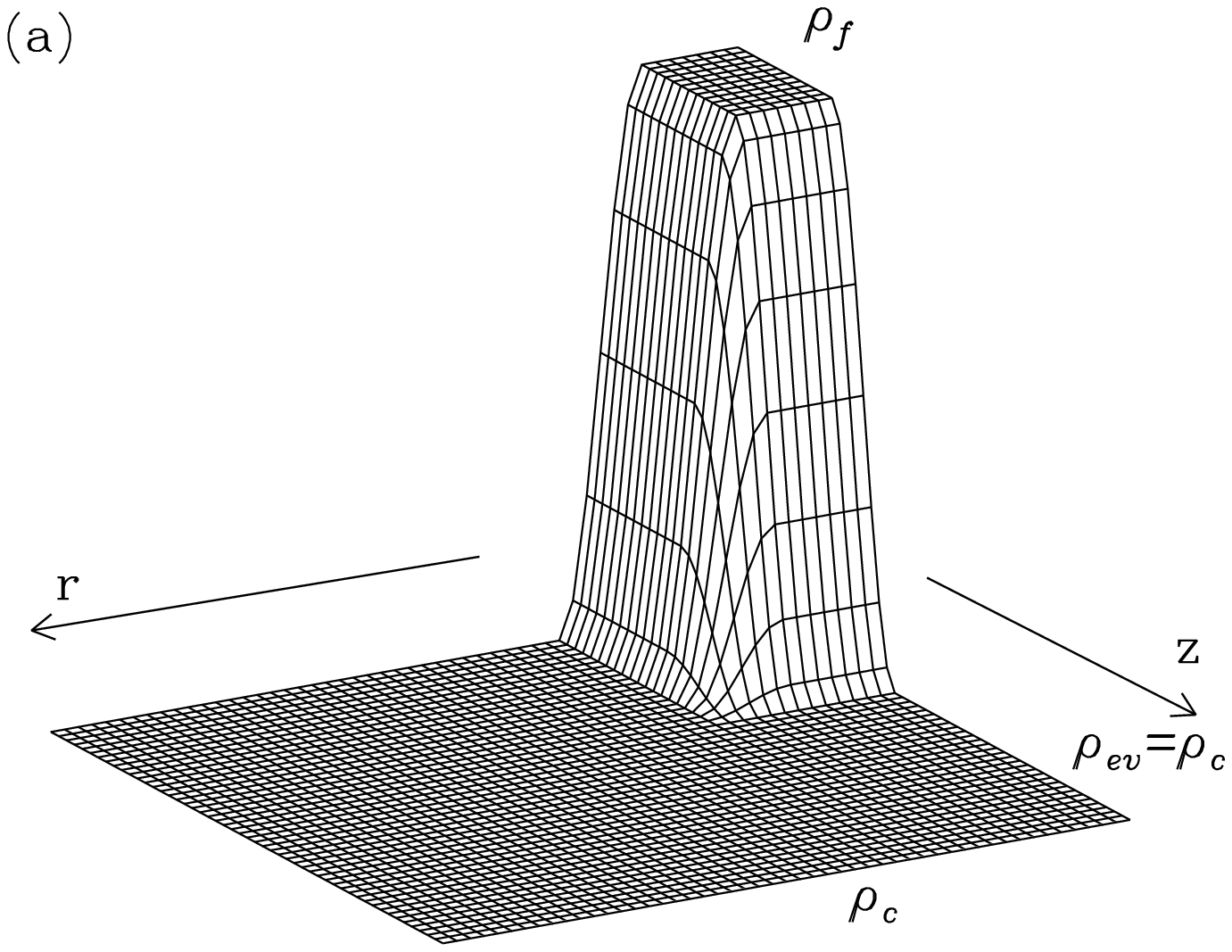}
  \includegraphics[width=5.8cm,height=4.55cm]{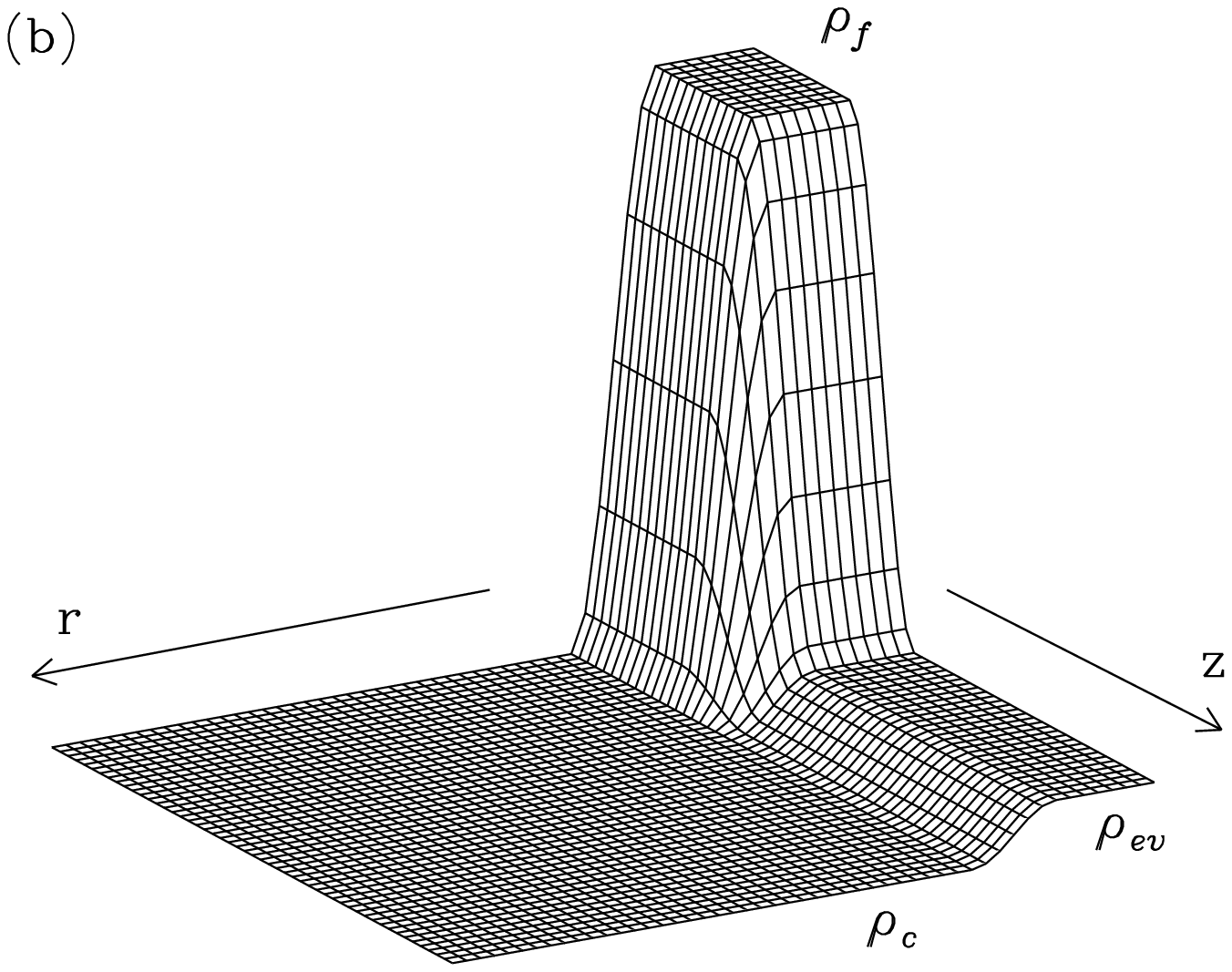}
   \includegraphics[width=5.8cm,height=4.55cm]{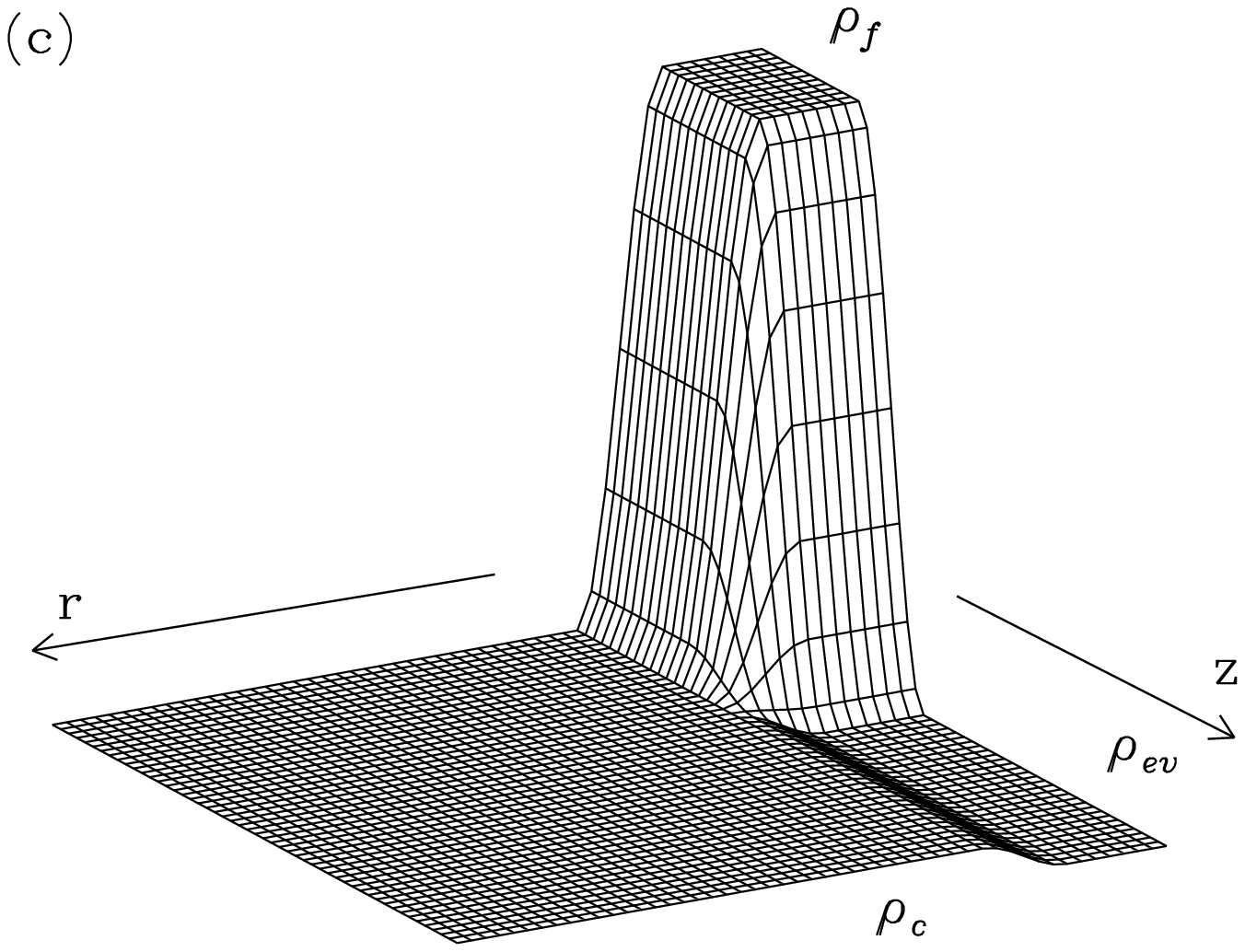}
 \caption{Density distribution in the ($r$, $z$)-plane in the domain $r\in[0,r_{\rm max}]$,  $z\in[0,L/2]$ for prominence thread models with non-uniform radial and longitudinal structuring. The threads are defined by a cool and dense part with density, $\rho_{\rm f}$ and length $L_{\rm thread}$ and an evacuated part of the tube with density $\rho_{\rm ev}$ and length $L-L_{\rm thread}$. These two regions connect non-uniformly to the coronal surrounding medium with density $\rho_{\rm c}$. Depending on the value of $\rho_{\rm ev}$, three different situations are possible: 
 (a) $\rho_{\rm ev}=\rho_{\rm c}$, (b) $\rho_{\rm c}\leq\rho_{\rm ev}\leq\rho_{\rm f}$, and (c) $\rho_{\rm ev}\leq\rho_{\rm c}$.\label{model}}
\end{figure*}

\section{Thread model, linear MHD wave equations, and numerical method}\label{sect2}

We consider an individual and isolated prominence thread in a gravity-free static equilibrium in the zero plasma-$\beta$ approximation. The fine structure 
is modelled by means of a cylindrically symmetric flux tube of radius $a$ and length $L$. In a system of cylindrical coordinates ($r$, $\varphi$, $z$) 
with the $z$-axis coinciding with the axis of the tube, the magnetic field is pointing in the $z$-direction and has a uniform field strength. Because of the 
assumed  zero-$\beta$ approximation the density profile can be chosen arbitrarily. The non-uniform thread is then modelled as a density
enhancement with a two-dimensional distribution of density, $\rho(r,z)$  (see Fig.~\ref{modelfig}).  The density distribution has two non-uniform layers, with length $l$ and $l_{\rm z}$ 
in the $r$- and $z$- directions, respectively.  The first is introduced so as to study the resonant damping of oscillations. The second produces irrelevant physical results, but enables 
us to avoid contact discontinuities and provides us with a continuous background density. Surface plots of the density distribution for different thread models are shown in Figure~\ref{model}, where we 
have made use of the symmetry of the system in the $r$- and $z$- directions and only plot their positive values. The dense part of the tube with 
prominence conditions, i.e., the thread, has a density $\rho_{\rm f}$ and occupies only part of the larger magnetic flux tube. It extends over a length 
$L_{\rm thread}$ in the $z$- direction. The rest of the internal part of the tube, with length $L-L_{\rm thread}$ in the longitudinal direction, is filled with 
plasma with a density $\rho_{\rm ev}$, with the subscript ``ev'' indicating the evacuated part of the tube.  All lengths are normalised by taking $a=1$. 
In contrast to \cite{soler102dthread}, the value of $\rho_{\rm ev}$ can be different from the coronal density, $\rho_{\rm c}$,  and may have values lower  
than $\rho_{\rm c}$ up to a value equal to the filament density, $\rho_{\rm f}$, case in which the full tube is occupied with dense cool plasma and 
we recover the one-dimensional case.  
Both regions along the axis of the tube are connected by means of a non-uniform transitional layer of length  $l_{\rm z}$ to produce a smooth 
longitudinal profile. The density variation at this layer, with  $L_{\rm thread}-l_{\rm z}/2\leq z \leq L_{\rm thread}+l_{\rm z}/2$, can be expressed as 

\begin{equation}
\rho_z(z)=\frac{\rho_{\mathrm f}}{2}\left[\left(1+\frac{\rho_{\mathrm {ev}}}{\rho_{\mathrm f}}\right)-\left(1-\frac{\rho_{\mathrm {ev}}}
{\rho_{\mathrm f}}\right)\sin{\frac{\pi(z-L_{\mathrm {thread}})}{l_{\mathrm  z}}}\right],
\end{equation}

\noindent
for $r\leq a-l/2$. 
As for the radial direction, the internal filament plasma, with density $\rho_{\rm f}$, is connected to the external coronal plasma, with density $\rho_c$, by 
means of a non-uniform transitional layer of thickness $l$, defined in the interval [$a-l/2$, $a+l/2$], that can vary in between $l/a=0$ (homogeneous tube) and $l/a=2$ (fully non-uniform tube). 
In contrast to the one-dimensional model used by \citet{Arregui08thread} the dense plasma of the thread does not occupy the full length of the tube, 
hence the radial variation of the plasma density for  $z\leq L_{\rm thread}-l_{\rm z}/2$ and $a-l/2\leq r \leq a+l/2$ is given by

\begin{equation}
\rho_r(r)=\frac{\rho_{\mathrm f}}{2}\left[\left(1+\frac{\rho_{\mathrm c}}{\rho_{\mathrm f}}\right)-\left(1-\frac{\rho_{\mathrm c}}
{\rho_{\mathrm f}}\right)\sin{\frac{\pi(r-a)}{l}}\right].
\end{equation}

\noindent
Radial non-uniformity is not restricted to the dense part, as in \citet{soler102dthread}, but can be present also in the evacuated part of the tube. As a consequence, there is an overlap region of radial and longitudinal non-uniform layers, for $a-l/2\leq r \leq a+l/2$ and $L_{\rm thread}-l_{\rm z}/2\leq z \leq L_{\rm thread}+l_{\rm z}/2$, where the density is given by

\begin{equation}
\rho_{rz}(r,z)=\frac{\rho_{\mathrm z}(z)}{2}\left[\left(1+\frac{\rho_{\mathrm c}}{\rho_{\mathrm z}(z)}\right)-\left(1-\frac{\rho_{\mathrm c}}
{\rho_{\mathrm z}(z)}\right)\sin{\frac{\pi(r-a)}{l}}\right].
\end{equation}

\noindent
In the evacuated part of the tube, for $z\geq L_{\rm thread}+l_{\rm z}/2$ and $a-l/2\leq r \leq a+l/2$, the radial density profile is given by

\begin{equation}\label{rhoprima}
\rho_r(r)=\frac{\rho_{\mathrm {ev}}}{2}\left[\left(1+\frac{\rho_{\mathrm c}}{\rho_{\mathrm {ev}}}\right)-\left(1-\frac{\rho_{\mathrm c}}
{\rho_{\mathrm {ev}}}\right)\sin{\frac{\pi(r-a)}{l}}\right].
\end{equation}

\noindent
Note that since the case $\rho_{\rm ev}\leq\rho_{\rm c}$ is not excluded, the slope of the radial density profile in the evacuated part of 
the tube can be positive (if $\rho_{\rm ev} < \rho_{\rm ev}$), negative (if $\rho_{\rm ev} > \rho_{\rm ev}$) or zero (if $\rho_{\rm ev}=\rho_{\rm ev}$). 
Finally, for $r\geq a+l/2$ we reach the coronal medium and $\rho(r,z)=\rho_{\rm c}$.  The model adopted in this paper removes unrealistic 
discontinuities in the density distribution, considered in previous works, and allows the theoretical modelling of a  large number of threads by 
simply considering different geometrical and physical parameter values for, e.g., the three different situations outlined in Figure~\ref{model}.

This paper is concerned with standing kink waves in prominence threads.
To study small amplitude thread oscillations, we consider the linear resistive MHD wave equations for perturbations of the form 
$f(r,z)\sim exp(i(\omega t+m\varphi))$ with constant resistivity, $\eta$. Here $m$ is the azimuthal wave-number and 
$\omega=\omega_{\rm R}+i\omega_{\rm I}$, the complex oscillatory frequency. For resonantly damped solutions, the real part of the frequency 
gives the period of the oscillation, $P=2\pi/\omega_{\rm R}$, while the imaginary part is related to the damping time, $\tau_d=1/\omega_{\rm I}$.  
Magnetic diffusion is only included here to avoid the singularity of the MHD equations at the resonant position, but diffusion has no effect 
on the resonant damping time-scales. Oscillations are then governed by the following set of partial differential equations for the two components of the 
velocity perturbation, $v_r$ and $v_\varphi$, and the three components of the perturbed magnetic field, $b_r$, $b_\varphi$, and $b_z$,

\begin{figure*}
\includegraphics[width=5.8cm,height=4.55cm]{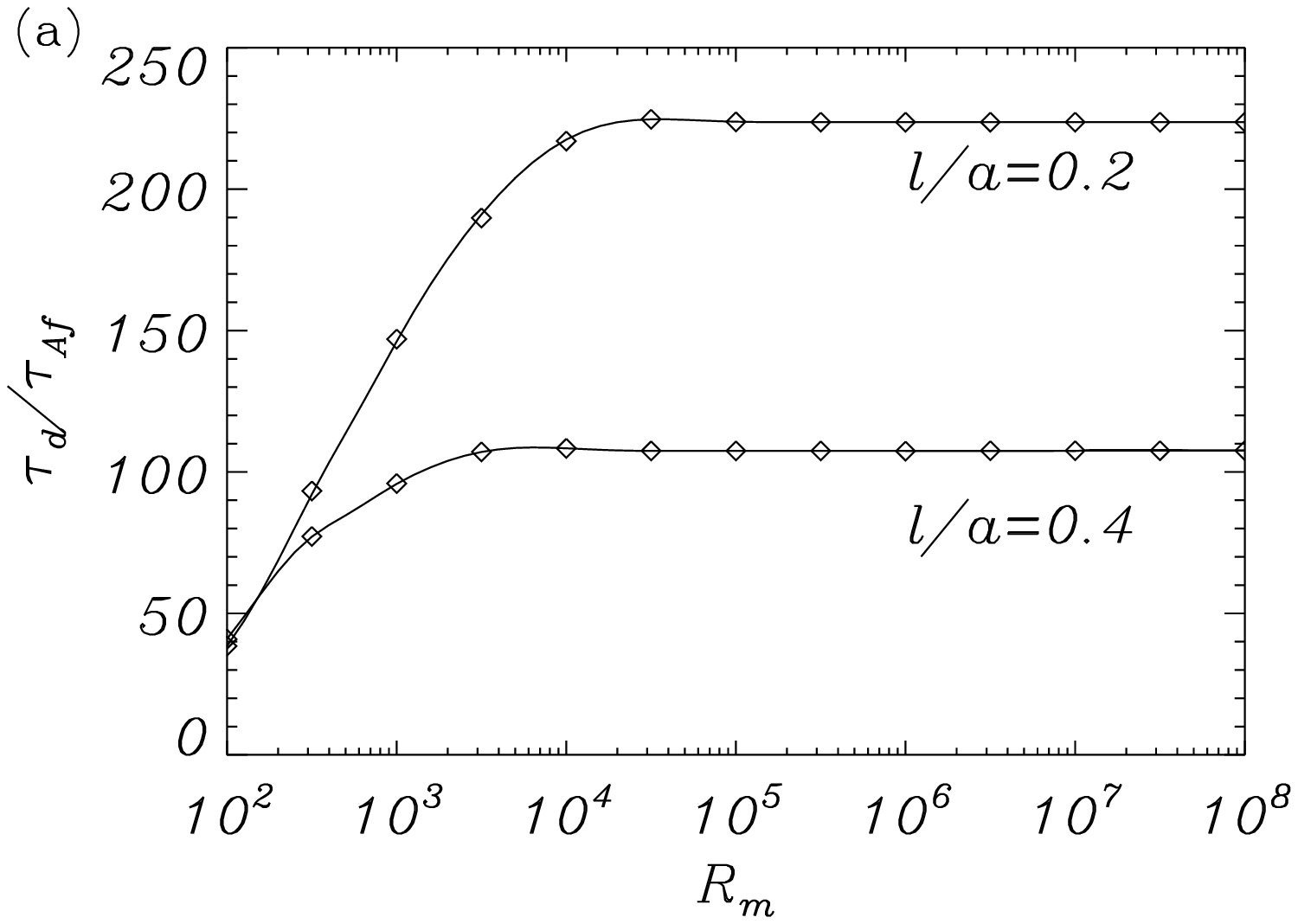}
 \includegraphics[width=5.8cm,height=4.55cm]{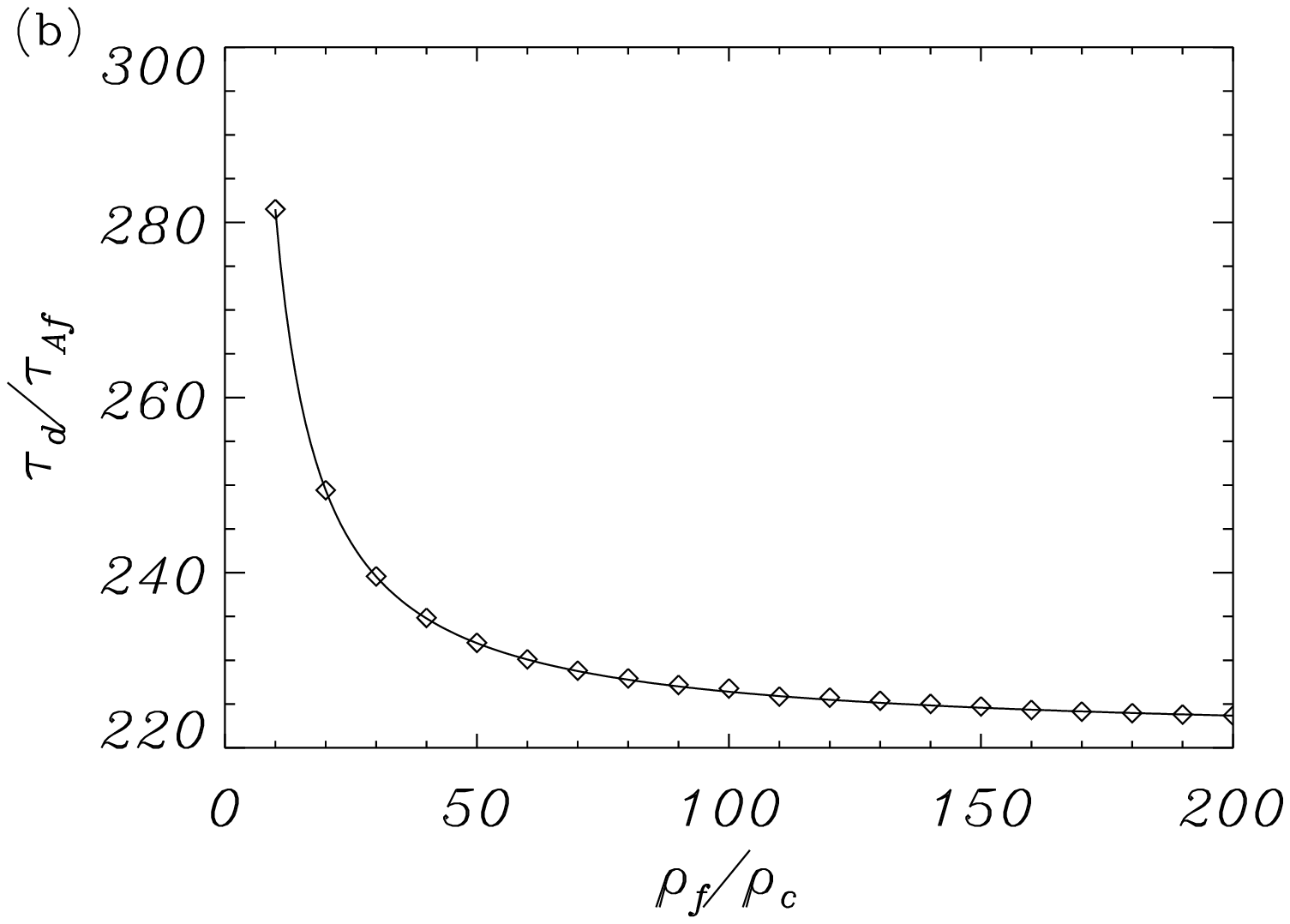}
 \includegraphics[width=5.8cm,height=4.55cm]{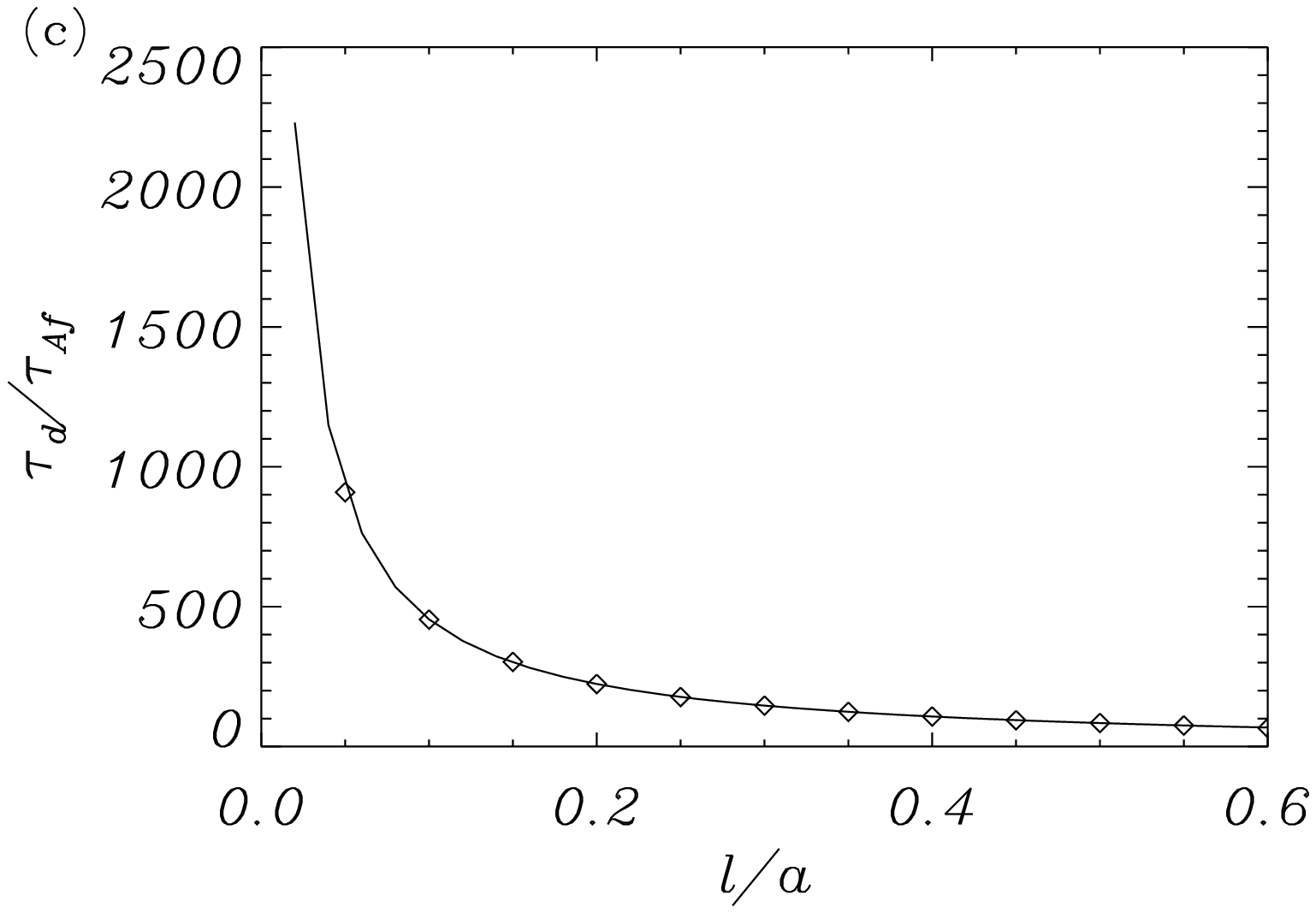}
 \caption{Damping time, in units of the internal filament Alfv\'en crossing time, $\tau_{\rm Af}=a/v_{\rm Af}$, as function of the three relevant parameters for kink oscillations in one-dimensional thread models: (a) the magnetic Reynolds number, $R_{\rm m}=v_{\rm Af}a/\eta$, (b) the density contrast, $\rho_{\rm f}/\rho_{\rm c}$, and (c) the transverse inhomogeneity length-scale, $l/a$. Solid lines indicate the 1D solution, while symbols represent the numerical 2D solutions obtained with $L_{\rm thread}=L$. In (a) and (c) the density contrast is
$\rho_{\rm f}/\rho_{\rm c}=200$. In (b), $l/a=0.2$.  In (b) and (c) $R_{\rm m}=10^6$. All computations have been performed in a two-dimensional grid with $N_r=401$ and $N_z=51$ points, with 250 grid-points in the radial transitional layer. Lengths are normalised to $a=1$ and $L=50a$. \label{check}}
\end{figure*}

\begin{eqnarray}
i\omega  v_r&=&\frac{B}{\mu\rho}\left(\frac{\partial b_r}{\partial z}-\frac{\partial b_z}{\partial r}\right),\label{first}\\
i\omega  v_\varphi&=&-\frac{B}{\mu\rho}\left(\frac{im}{r} b_z-\frac{\partial b_\varphi}{\partial z}\right),\\
i\omega  b_r&=&B \frac{\partial v_r}{\partial z}+\eta\left[\frac{\partial^2 b_r}{\partial r^2}-\frac{m^2}{r^2}b_r
+\frac{\partial^2 b_r}{\partial z^2}+\frac{1}{r}\frac{\partial b_r}{\partial r}\right.\nonumber\\
&-& \left. 2\frac{im}{r^2}b_{\varphi}-\frac{b_r}{r^2}\right], \\
i\omega b_\varphi&=&B\frac{\partial v_\varphi}{\partial z}+\eta\left[\frac{\partial^2 b_\varphi}{\partial r^2}-\frac{m^2}{r^2}b_\varphi+
\frac{\partial^2 b_\varphi}{\partial z^2}+\frac{1}{r}\frac{\partial b_\varphi}{\partial r}\right.\nonumber\\
&+& \left. 2\frac{im}{r^2}b_{r}-\frac{b_\varphi}{r^2}\right], \\
i\omega b_z&=&-B\left(\frac{\partial v_r}{\partial r}+\frac{v_r}{r}+\frac{im}{r}v_\varphi\right)+\eta\left[\frac{\partial^2 b_z}{\partial r^2}-
\frac{m^2}{r^2}b_z\right.\nonumber\\
&+& \left.\frac{\partial^2 b_z}{\partial z^2}+\frac{1}{r}\frac{\partial b_z}{\partial r}\right] . \label{last}
\end{eqnarray}

\noindent
Equations (\ref{first})--(\ref{last}), together with the appropriate boundary conditions, define an eigenvalue problem for resonantly 
damped modes. As the plasma-$\beta$=$0$, the slow mode is absent and there are no motions parallel to the equilibrium magnetic 
field, $v_z=0$. We further concentrate on perturbations with $m=1$, which represent kink waves that produce the transverse 
displacement of the axis of the tube. The MHD kink wave represents a wave mode with mixed fast and Alfv\'en character, its 
Alfv\'enic nature being dominant in and around the resonant position \citep{Goossens09}. The problem can be further simplified by 
making use of the divergence-free condition for the perturbed magnetic field

\begin{equation}
\frac{1}{r}\frac{\partial(r b_r)}{\partial r}+\frac{1}{r}\frac{\partial b_\varphi}{\partial \varphi}+\frac{\partial b_z}{\partial z}=0,\label{divb}
\end{equation}

\noindent
which reduces the system of equations to be solved to four, upon expressing $b_\varphi$ in terms of $b_r$ and $b_z$. Solutions 
to these equations are obtained by performing a normal mode analysis. Because of the complexity of the problem when a two-dimensional 
density $\rho(r,z)$ is considered,  numerical solutions to the frequency and spatial structure of eigenfunctions in the ($r,z$)- plane are 
computed  using PDE2D \citep{sewell05}, a general-purpose partial differential equation solver. The code uses finite elements and allows 
the use of non-uniformly distributed grids, which are needed to properly resolve the large gradients that arise in the vicinity of resonant positions. 
Different grid resolutions have been tested so as to assure the proper computation of the resonant eigenfunctions. Magnetic dissipation has no 
influence on the damping of kink modes due to resonant absorption, a condition that has to be checked in all numerical computations, but allows 
us to properly compute the spatial distribution of perturbations in the resonance.  We have made use of the symmetry of the system and 
solutions are computed in the domain $r\in[0, r_{\rm max}], z\in[0, L/2]$. Non-uniform grids are used in both directions, to 
properly resolve the regions with $r\in[a-l/2,a+l/2]$ and $z\in[L_{\rm thread}/2-l_{\rm z}/2,L_{\rm thread}/2+l_{\rm z}/2]$. The first is located so 
as to include the non-uniform transitional layer in the radial direction, while the second embraces the non-uniform transitional layer along the tube. 
As for the boundary conditions, we equate them to the spatial distribution of perturbations for the fundamental kink mode. The two perturbed 
velocity components, $v_r$ and $v_\varphi$ and the compressive component of the perturbed magnetic field, $b_z$ have vanishing longitudinal 
derivatives at $z=0$, which corresponds to the apex of the flux tube, while they vanish at $z=L/2$, because of the line-tying boundary condition at 
the photosphere. In the radial direction, they also have vanishing radial derivative at the axis of the tube, $r=0$, while we impose the vanishing of 
the perturbed velocity components far away from the tube in the radial direction, hence ($v_r$, $v_\varphi$) $\rightarrow 0$ as $r\rightarrow\infty$, 
a condition that is accomplished by setting the perturbations equal to zero at $r=r_{\rm max}$, where $r_{\rm max}$, the upper limit of the domain in 
the radial direction, has to be chosen to be sufficiently far to properly compute the drop-off rate of perturbations in the radial direction. 
In all our computations, we have considered $r_{\rm max}=20a$.

\begin{figure*}
 \includegraphics[width=5.8cm,height=4.55cm]{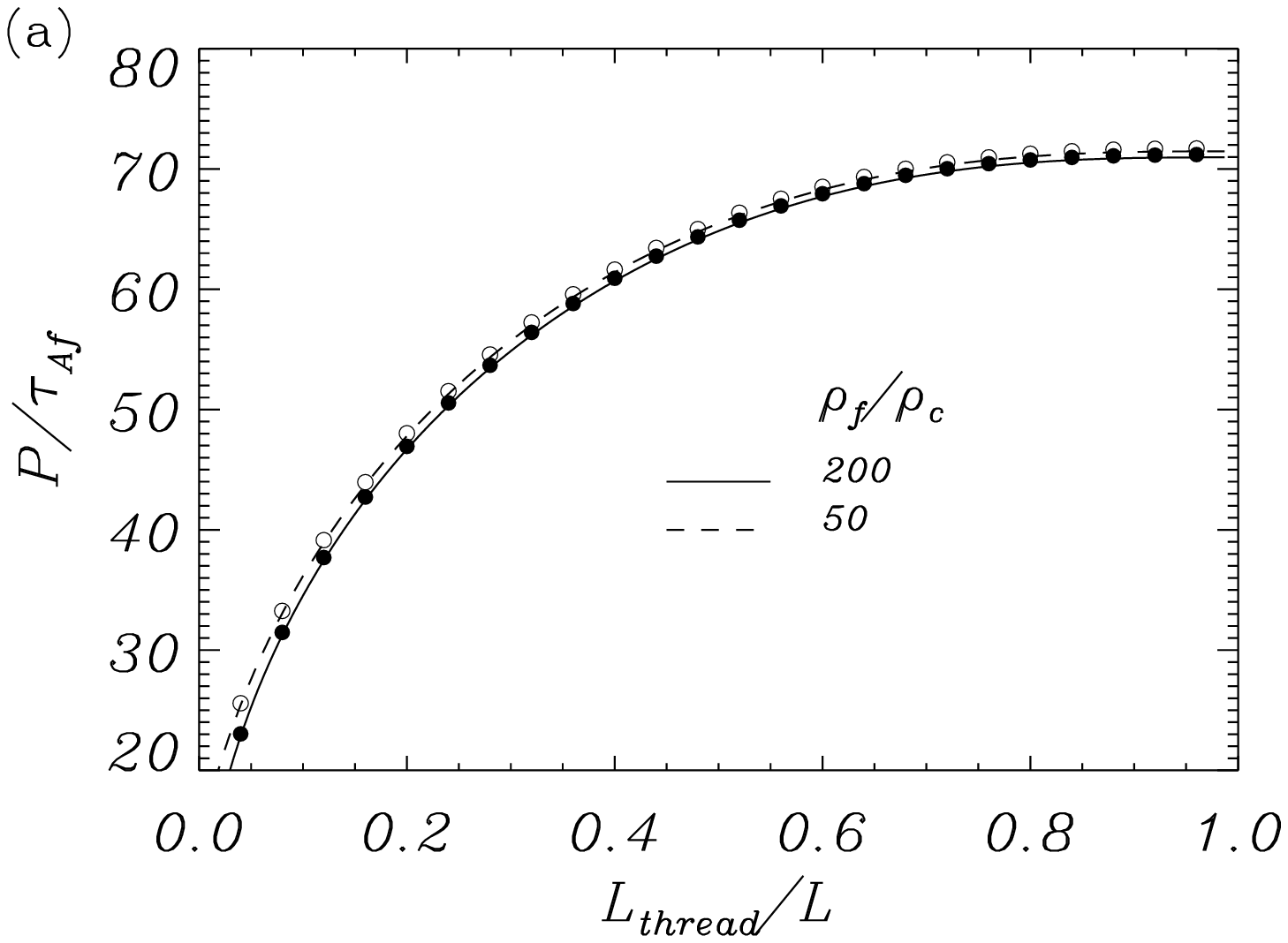}
 \includegraphics[width=5.8cm,height=4.55cm]{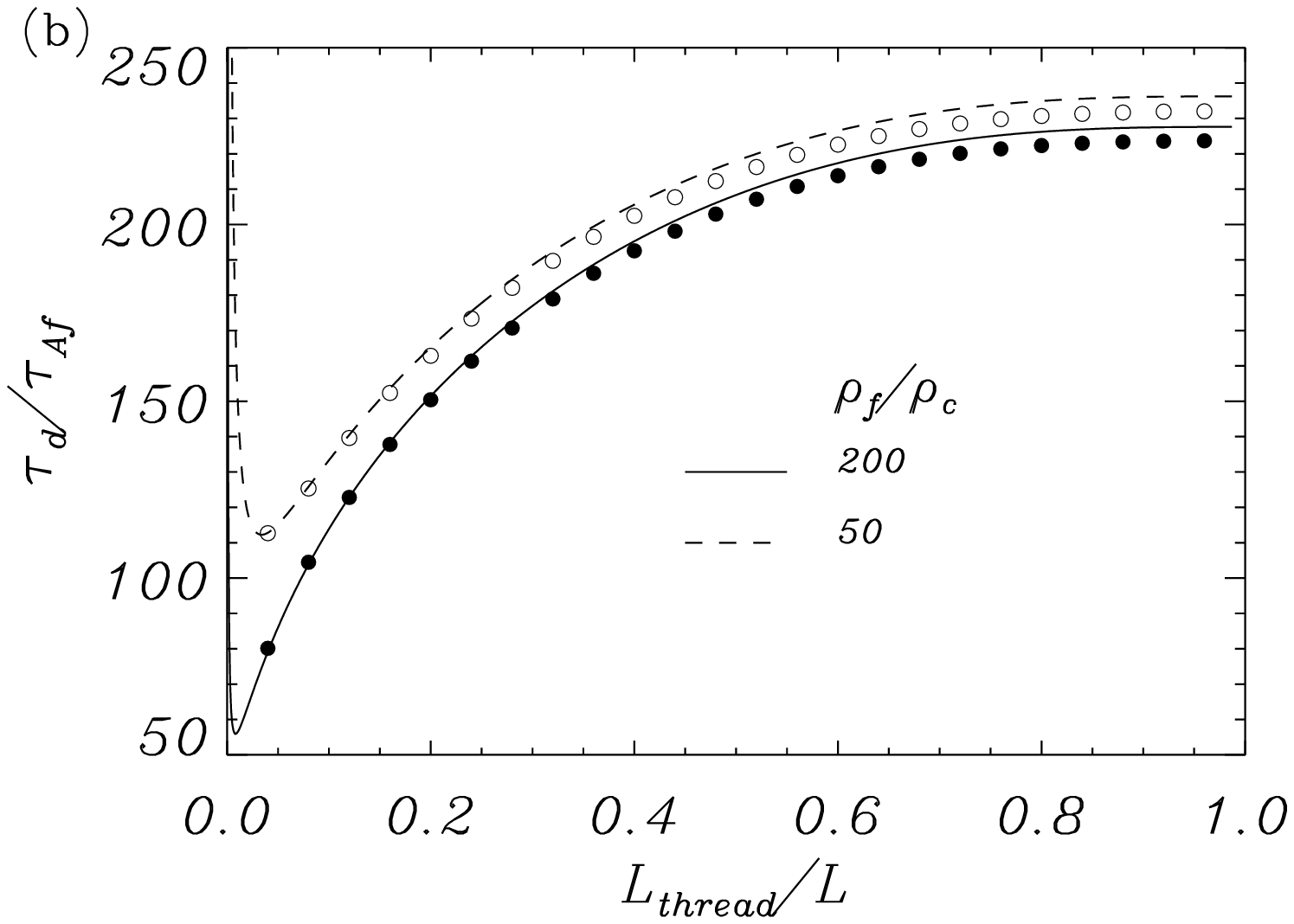}
 \includegraphics[width=5.8cm,height=4.55cm]{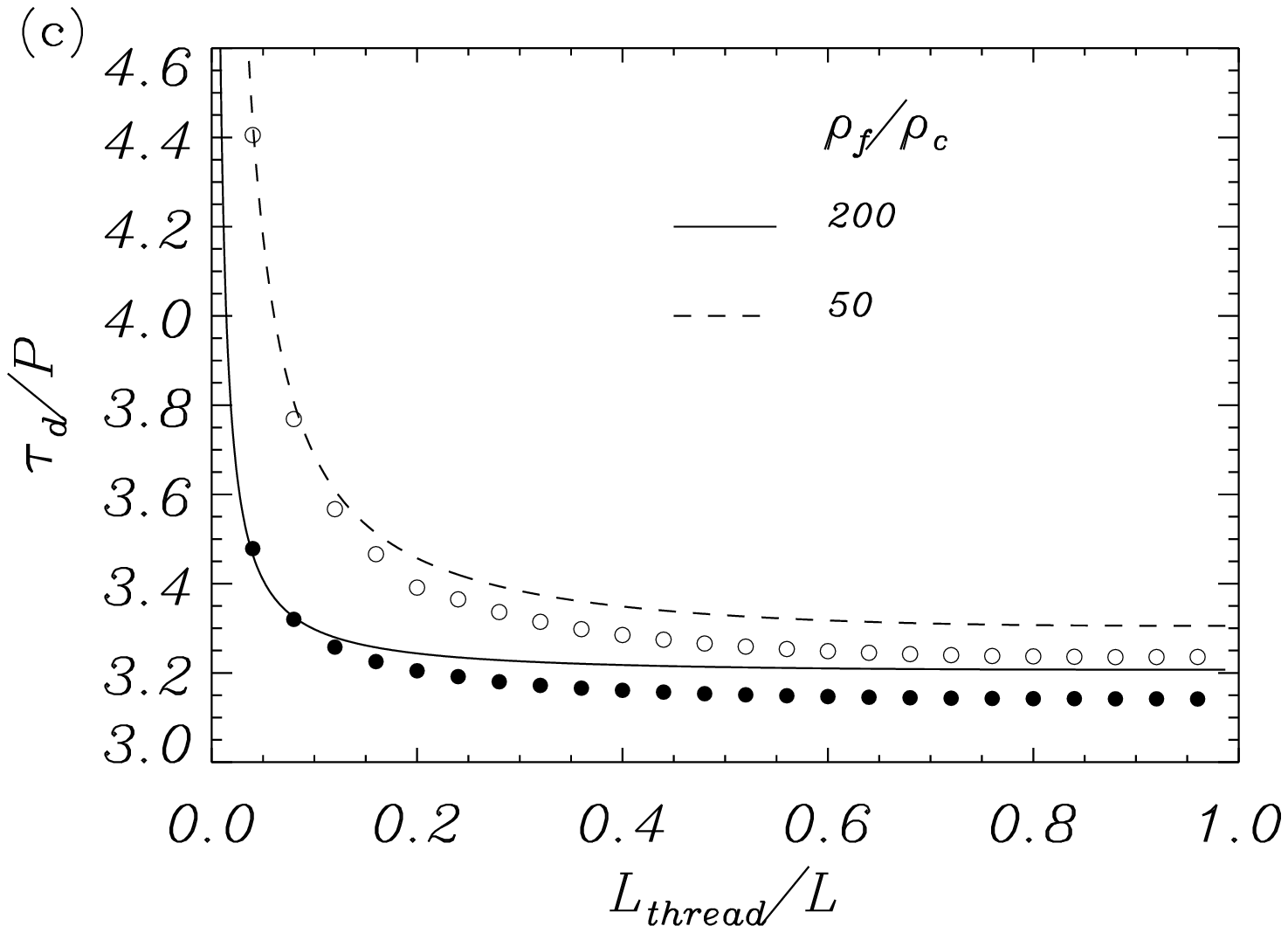}\\
 \includegraphics[width=5.8cm,height=4.55cm]{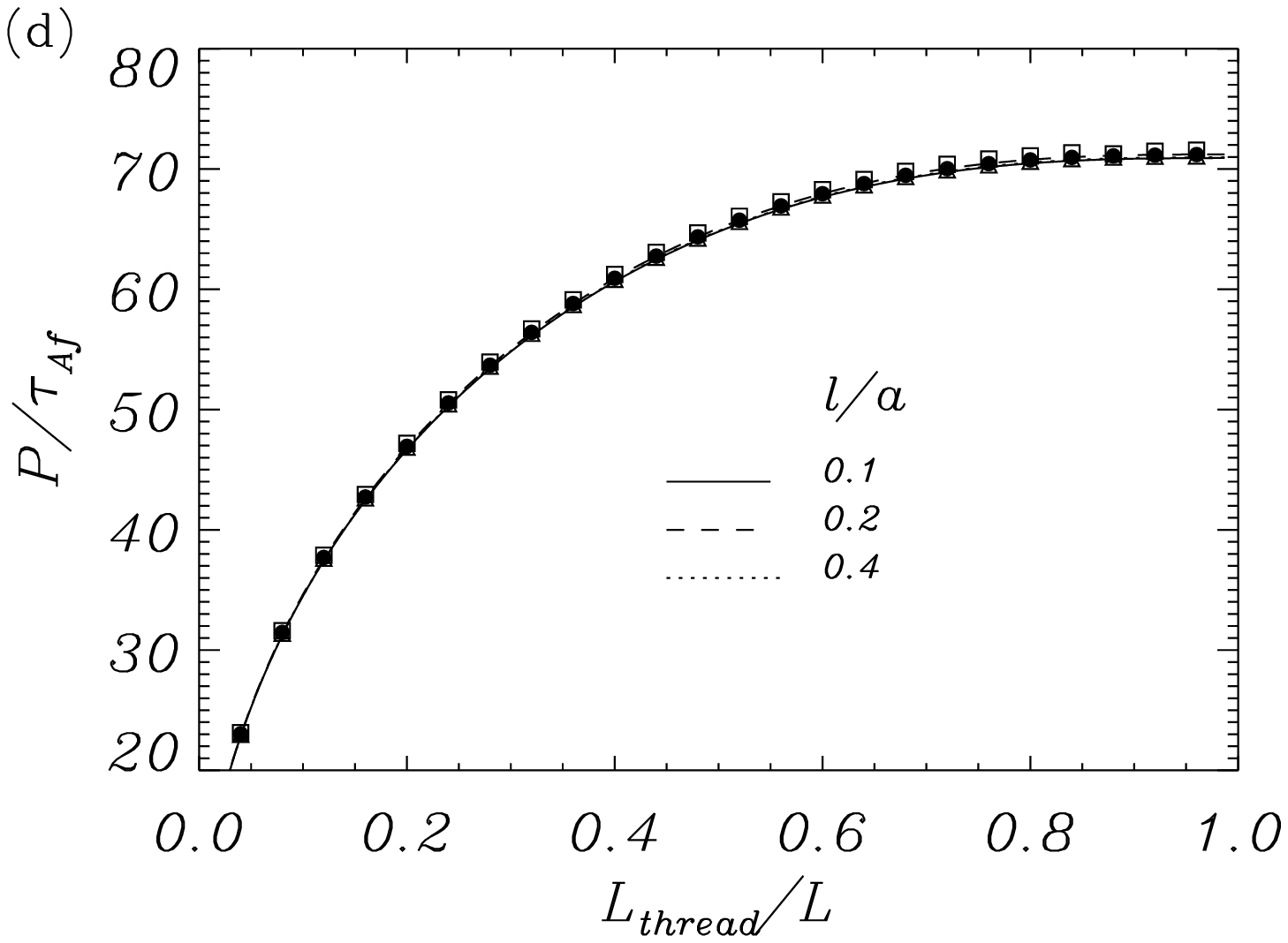}
 \includegraphics[width=5.8cm,height=4.55cm]{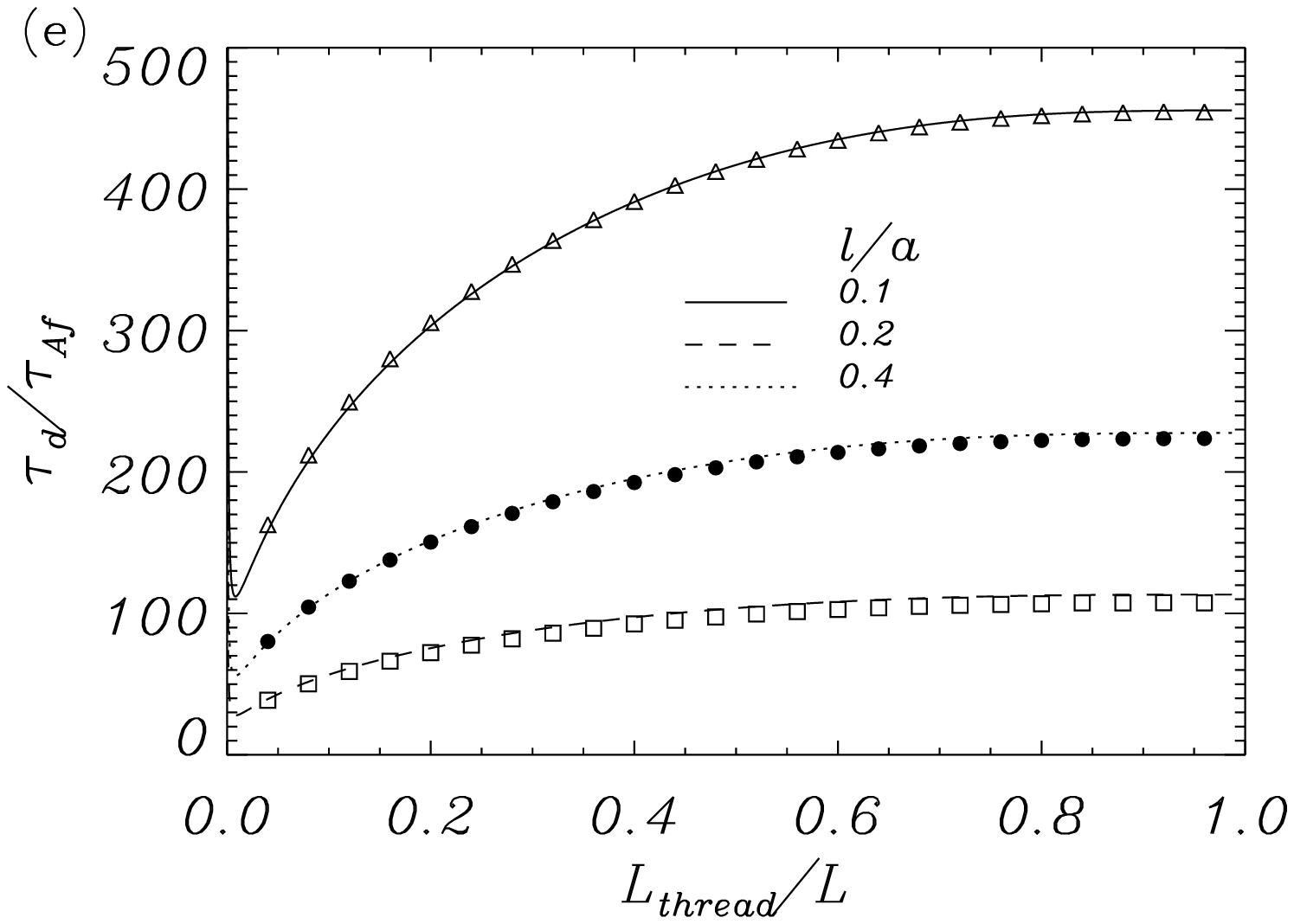}
 \includegraphics[width=5.8cm,height=4.55cm]{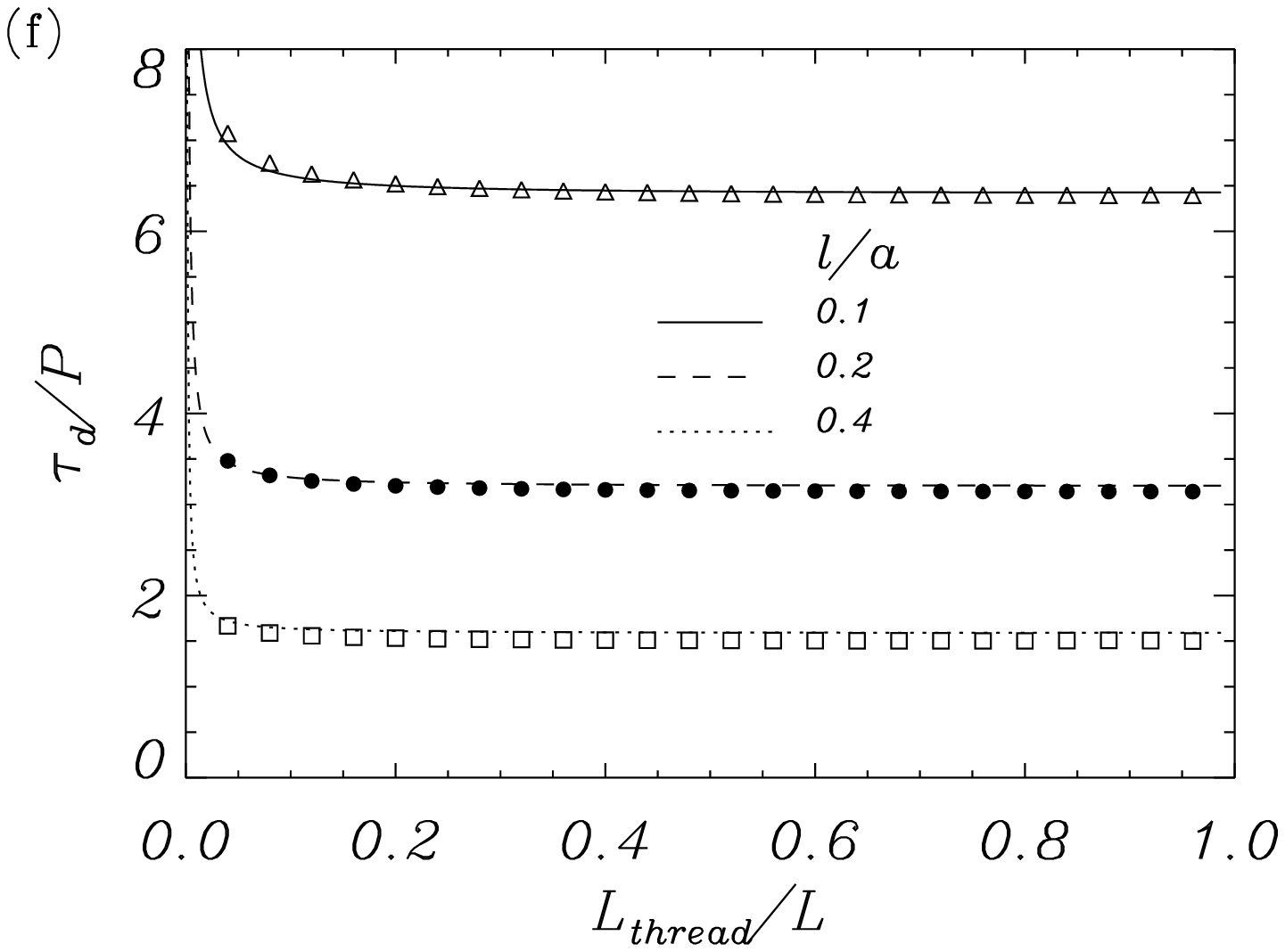}

\caption{(a)-(c): period, damping time, and damping ratio as a function of the length of the thread for thread models with $l/a=0.2$ and for two values of the density contrast. (d)-(f): the same quantities for thread models with $\rho_{\rm f}/\rho_{\rm c}=200$ and for three values of the transverse inhomogeneity length-scale.  In all figures $a=1$, $R_{\rm m}=10^6$, $\rho_{\rm ev}=\rho_{\rm c}$, and $l_{\rm z}=a$. Times are shown in units of the internal filament Alfv\'en crossing time, $\tau_{\rm Af}=a/v_{\rm Af}$. Symbols correspond to fully numerical 2D computations  while the different line styles represent the approximate 2D solutions obtained by \citet{soler102dthread}. All computations have been performed in a two-dimensional grid with $N_r=401$ and $N_z=51$ points, with 250 grid-points in the resonant layer. Lengths are normalised to $a=1$ and $L=50a$. \label{figlthread}}
\end{figure*}

\section{Analysis and results}\label{sect3}

\citet{Arregui08thread}, in their analysis of the damping of kink oscillations in one-dimensional thread models, showed that the parameters that 
determine the temporal attenuation of oscillations are the density contrast, $\rho_{\rm f}/\rho_{\rm c}$ and the width of the non-uniform transitional 
layer, $l/a$. The damping ratio is rather dependent on the first parameter for low values of the density contrast, but stops being dependent in the 
high contrast ratio regime, typical of prominence plasmas. The strongest influence comes from the width of  the transitional layer, with the damping time rapidly decreasing for increasing values of $l/a$.  Before we deal with the additional parameters introduced by the longitudinal density structuring in  
two-dimensional thread models, solutions to Equations~(\ref{first})--(\ref{last}) have been first obtained by using  a two-dimensional (2D)  density 
distribution with $L_{\rm thread}=L$. The purpose of these numerical experiments has been to check the correct behaviour of the code by reproducing 
the results obtained by \citet{Arregui08thread}. In addition, we have also considered the magnetic Reynolds number, $R_{\rm m}=v_{\rm Af}a/\eta$, 
which should not affect the computed damping times, in the limit of large Reynolds numbers. Figure~\ref{check} displays the obtained results.  The 
damping time of resonantly damped kink waves is independent of the magnetic Reynolds number, as long as this quantity is large enough for 
resonance absorption to be the operating  damping mechanism. This regime (see the plateau regions) is obtained for different values of $R_{\rm m}$ 
when different transitional layers are considered. Figure~\ref{check}a shows a perfect agreement between the 1D results  and the current computations 
using the 2D code. The perfect correspondence between 1D and 2D computations for the damping time as a function of the density contrast in the radial 
direction is shown in Figure~\ref{check}b. Finally, the most important parameter that determines the damping of transverse thread oscillations is the 
width of the non-uniform transitional layer in the radial direction. The damping time strongly decreases when this parameter is increased, as can be 
seen in Figure~\ref{check}c, which again shows a very good agreement between the values computed in 2D  and the previous 1D computations. 
These results were in agreement with previous works in the context of the damping of coronal loop transverse oscillations 
\citep[e.g.][]{goossens92,RR02,GAA02}

Once we are confident about the goodness of the code, we consider the new ingredients introduced by the two-dimensional nature of the 
prominence thread models considered in this work. These new ingredients are the length of the thread, $L_{\rm thread}$, i.e., the length of the part 
of the magnetic flux tube filled with dense absorbing plasma, and the density in the evacuated part of the tube, $\rho_{\rm ev}$.
The non-uniform transitional layer in density between both regions along the tube, $l_{\rm z}$, has an irrelevant effect on the oscillatory period for low harmonics in the 
longitudinal direction, as shown by  \citet{diazmargarita08}. Our computations (not shown here) confirm the finding by \citet{diazmargarita08} on the negligible importance of this 
parameter for the oscillatory period and show a similar irrelevance concerning the damping time by resonant absorption. For this reason, 
we have concentrated our analysis on the remaining two parameters, $L_{\rm thread}$ and $\rho_{\rm ev}$.

 \begin{figure*}
 \includegraphics[width=5.8cm,height=4.55cm]{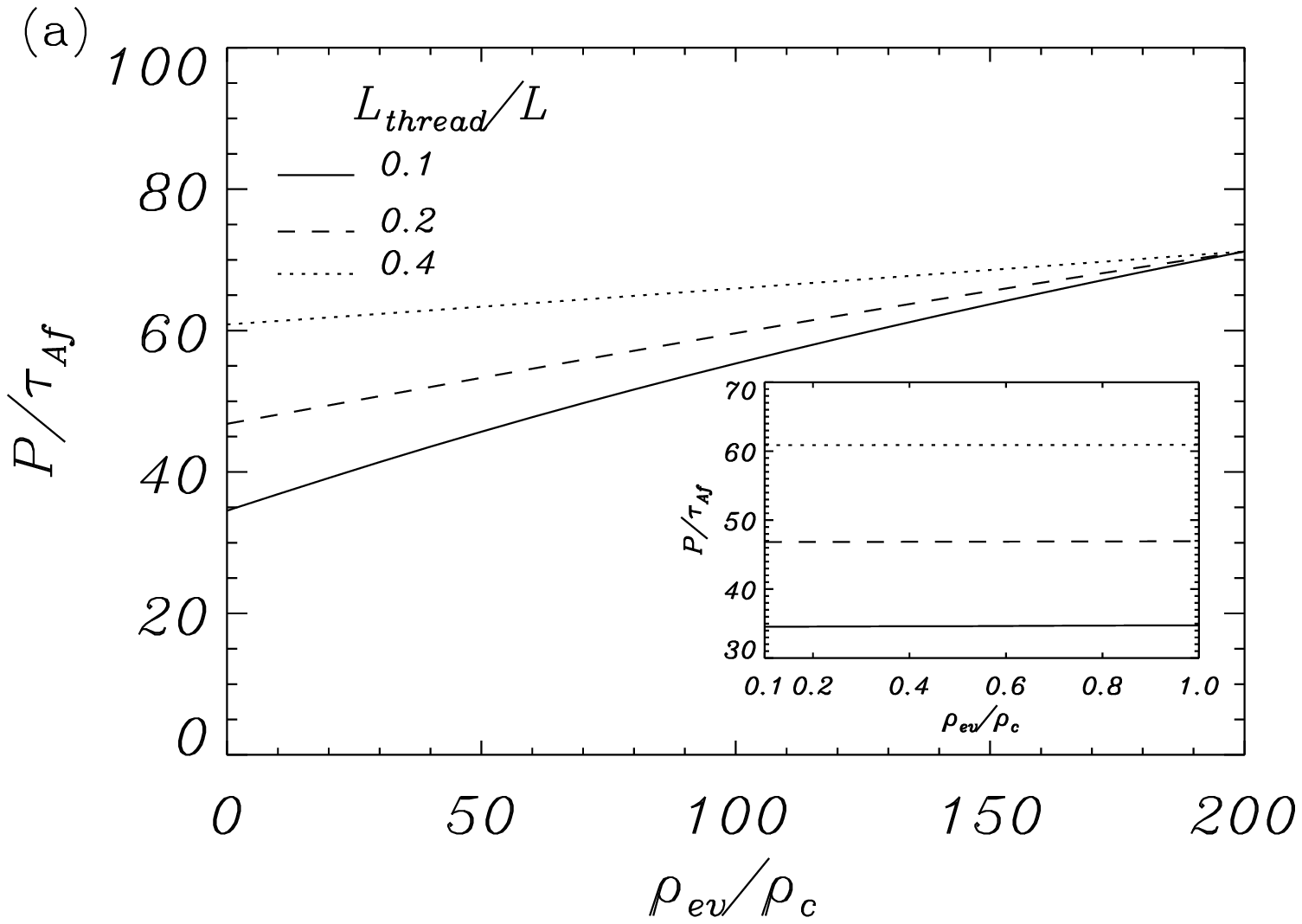}
 \includegraphics[width=5.8cm,height=4.55cm]{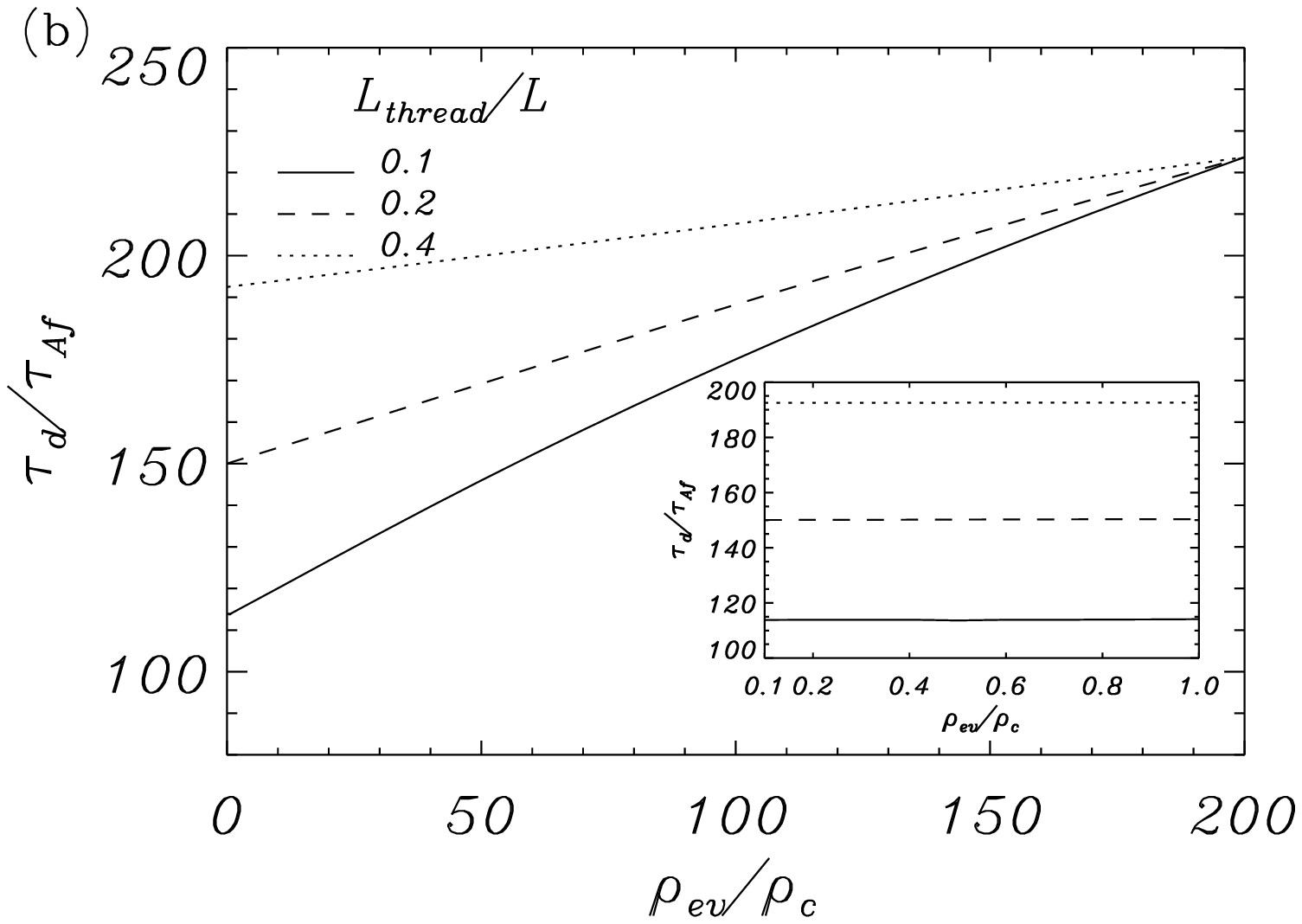}
  \includegraphics[width=5.8cm,height=4.55cm]{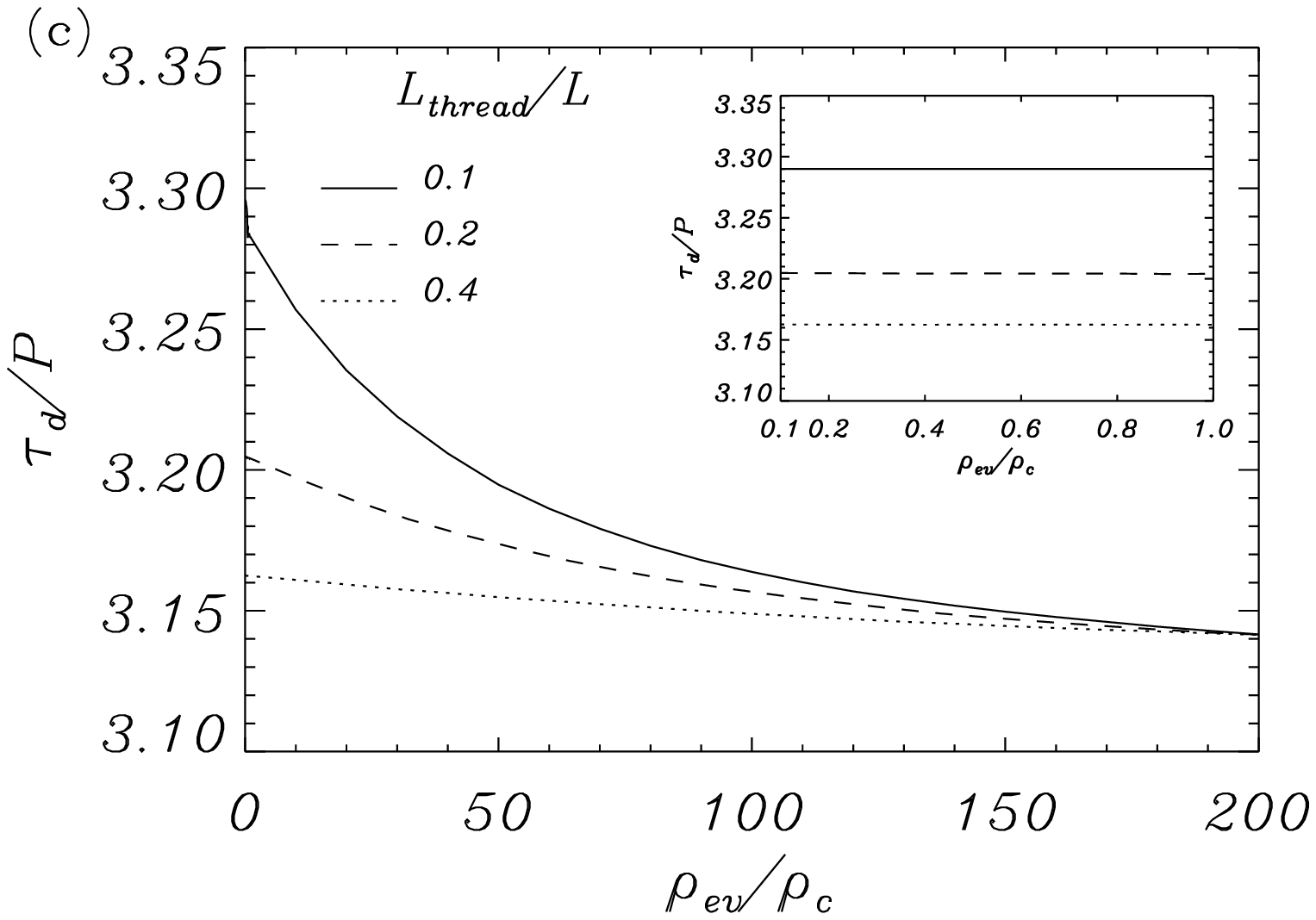}
\caption{Period, damping time, and damping ratio as a function of the density in the evacuated part of the tube for prominence threads with $\rho_{\rm f}/\rho_{\rm c}=200$, $a=1$, $R_{\rm m}=10^6$, $l_{\rm z}=a$, and $l/a=0.2$ and three values of the length of the thread. The main plots are for $\rho_{\rm c} \leq \rho_{\rm ev} \leq \rho_{\rm f}$, while the inset plots correspond to the range $0.1\rho_{\rm c} \leq \rho_{\rm ev} \leq \rho_{\rm c}$.
Times are shown in units of the internal filament Alfv\'en crossing time, 
$\tau_{\rm Af}=a/v_{\rm Af}$. All computations have been performed in a two-dimensional grid with $N_r=401$ and $N_z=51$ points, with 250 grid-points in the resonant layer. Lengths are normalised to $a=1$ and $L=50a$. \label{figrhoev}}
\end{figure*}

\subsection{Periods and damping times}\label{freqs}

We first consider the influence of the length of the thread on the period and damping of resonantly damped transverse thread oscillations. An 
initial analysis on this subject was presented by \cite{soler102dthread}, who considered the thin tube and thin boundary approximations 
(TTTB) in a two-dimensional thread model with transverse inhomogeneity only in the dense part of the tube. Our analysis goes beyond the TTTB approximations by considering a fully inhomogeneous two-dimensional density distribution and combining its influence with that of the density in the evacuated part of the tube, $\rho_{\rm ev}$.
 
We have first analysed the variation of period, damping time, and damping ratio, $\tau_d/P$, by setting $\rho_{\rm ev}=\rho_c$,  so that we 
mimic the case studied by \citet{soler102dthread} analytically. We start with a fully filled tube and gradually decrease the length of the thread. The obtained results, for two values of the density contrast between the filament and coronal plasma, are shown in Figures~\ref{figlthread}a-c. The period is strongly dependent on the length of the thread. It decreases by almost a factor of two when going from $L_{\rm thread}=L$ to $L_{\rm thread}=0.1L$. Figure~\ref{figlthread}a also shows that the oscillatory period is almost independent of the density contrast, once this parameter is large enough. 
As for standing kink waves in one-dimensional thread models \citep{Arregui08thread}, the kink frequency is a weighted mean of the internal and external Alfv\'en frequencies. Regardless of the density contrast, the period is allowed to vary in a a narrow range determined by a factor that goes from $\sqrt{2}$ to $1$, when going from $\rho_{\rm f}/\rho_{\rm c}=1$ to $\rho_{\rm f}/\rho_{\rm c}\rightarrow\infty$. For typical density contrasts in prominence plasmas, the period can be considered independent of the density contrast. The damping time produced by 
resonant absorption (Fig.~\ref{figlthread}b)  also decreases remarkably when the length of the cool and dense part of the tube is 
decreased. The decrease is also around a factor of two in the considered range of values for $L_{\rm thread}$. \citet{soler102dthread} find that in the 
TTTB limit the dependence of the period and the damping time with the length of the thread is exactly the same, hence any influence on the 
damping ratio, $\tau_{\rm d}/P$, is cancelled out. Outside the TTTB approximations, we find that this is not the case (see Fig.~\ref{figlthread}c), 
although the damping ratio is almost independent of the length of thread and only for very short threads a slight increase in the damping ratio is 
found when further decreasing this parameter. In Figure~\ref{figlthread} we overplot results obtained by \citet{soler102dthread}  by solving their dispersion 
relation. We see that there is a very good agreement and the differences, that are due to the simplifying assumptions of the analytical treatment, are rather small. One can observe an anomalous behaviour on the damping time computed by \cite{soler102dthread}, for small values of $L_{\rm thread}$. This is due to the simplifying assumptions considered to obtain the semi-analytic solution, that might not be entirely valid outside the long-wavelength limit.  Overall, our results confirm the validity of the analytical approximations obtained by \cite{soler102dthread} concerning the influence of the length of the thread on periods and damping times. In physical terms, the shortening of the length of the thread produces shorter period oscillations, since the physical system is equivalent to 
a fully filled tube, with the wavelength of oscillations replaced by a shorter effective wavelength. A physical explanation of the damping time dependence on the length of the thread is provided, by using energy arguments, in Sect.~\ref{energyanal}.

Similar conclusions can be extracted from the computations we have performed by fixing the density contrast and for three different values of the
width of the inhomogeneous layer. Figures~\ref{figlthread}d-f show the obtained results. They clearly show how strongly  the damping time 
and the damping ratio are influenced by the width of the transitional layer, also in 2D models, while the period of the oscillations is almost 
unaffected by the value of $l/a$. In view of the results displayed in Figure~\ref{figlthread}, we conclude that the length of the thread is a very 
important parameter. When allowing to vary from the limit of fully filled tube to 10\% filled tube, periods and damping times are decreased as much 
as 50\% percent. This is very relevant in connection to prominence seismology. We must note that, in principle, the length of the thread can be 
estimated directly from observations. However, the length of the supporting magnetic flux tube is much more difficult to estimate, since its end 
points are usually unobservable.

Our general density model enable us to analyse the effect of the density in the evacuated part of the tube on the oscillatory properties, as well.
The study of the influence of this parameter was not undertaken by \citet{soler102dthread}, and requires a fully numerical approach. We allow for $\rho_{\rm ev}$ to be 
different from the coronal density, $\rho_{\rm c}$. When changing the density in the evacuated part of the tube, in addition to the radial non-uniform layer that 
connects the prominence material to the corona we are introducing an additional radial non-uniform layer in between the evacuated part of the tube and the corona. The density profile in this layer and its slope depend on the relative values of $\rho_{\rm ev}$ and $\rho_{\rm c}$, as given in Equation~(\ref{rhoprima}) and shown in Figs.~\ref{model}b and c. Instead of analysing the effect of $\rho_{\rm ev}$ on the oscillatory properties in a separate manner, we have selected three representative values for the length of the thread and have computed periods, damping times, and damping ratios as a  function of the density in the evacuated part of the tube, measured in units of the coronal density.  We have split our analysis into two parts.

\begin{figure*}
 \includegraphics[width=5.8cm,height=4.55cm]{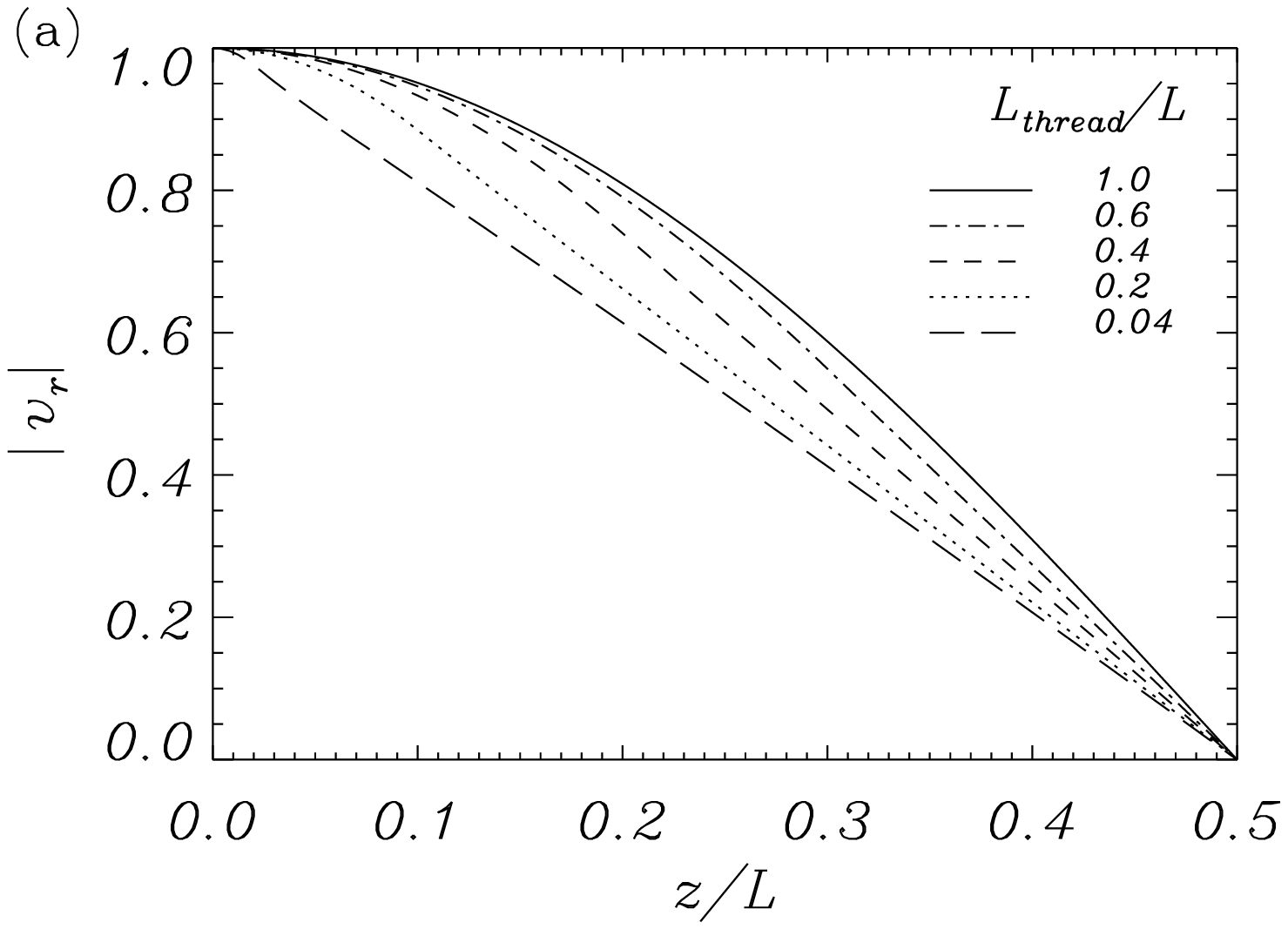}
 \includegraphics[width=5.8cm,height=4.55cm]{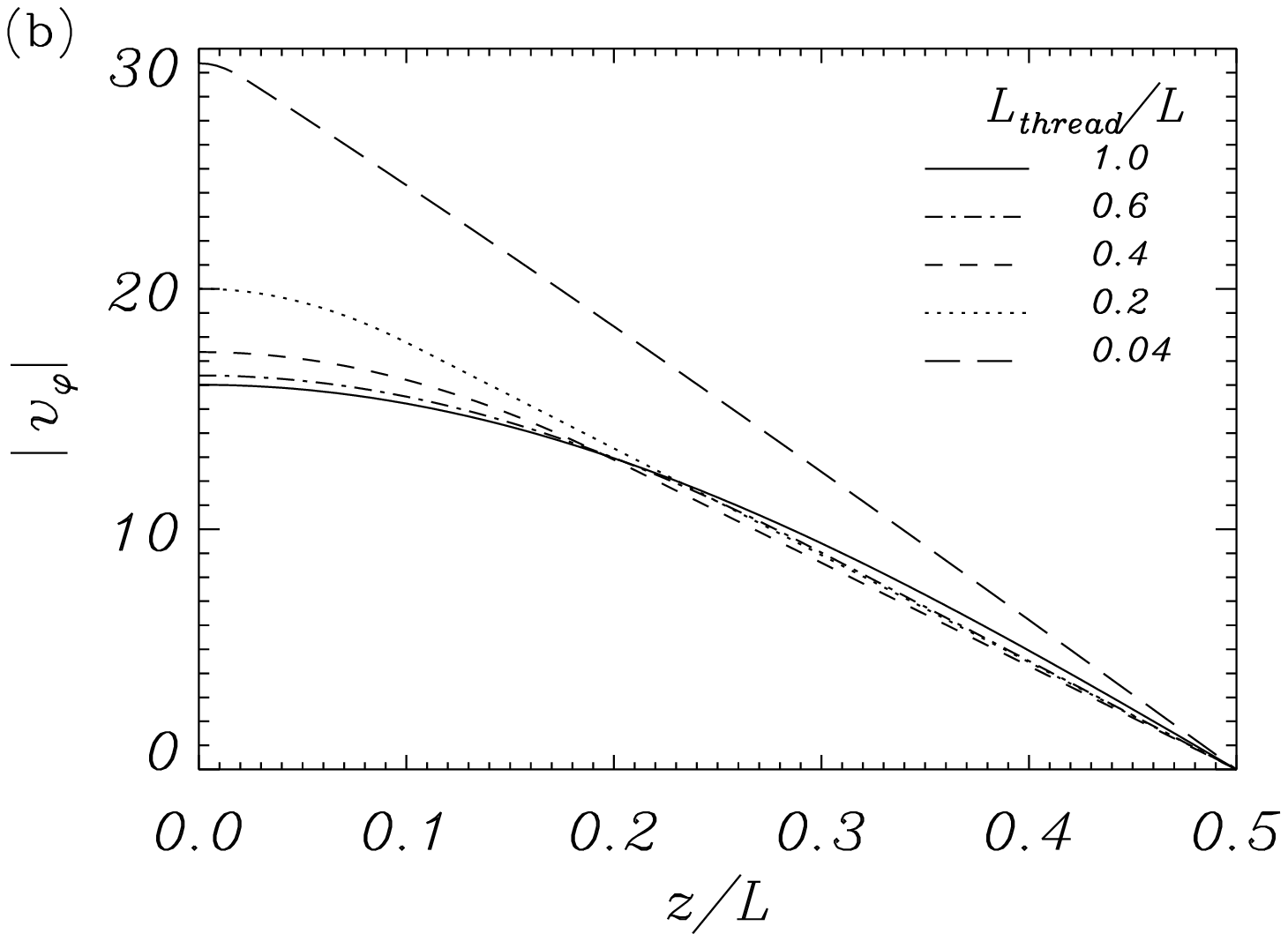}
 \includegraphics[width=5.8cm,height=4.55cm]{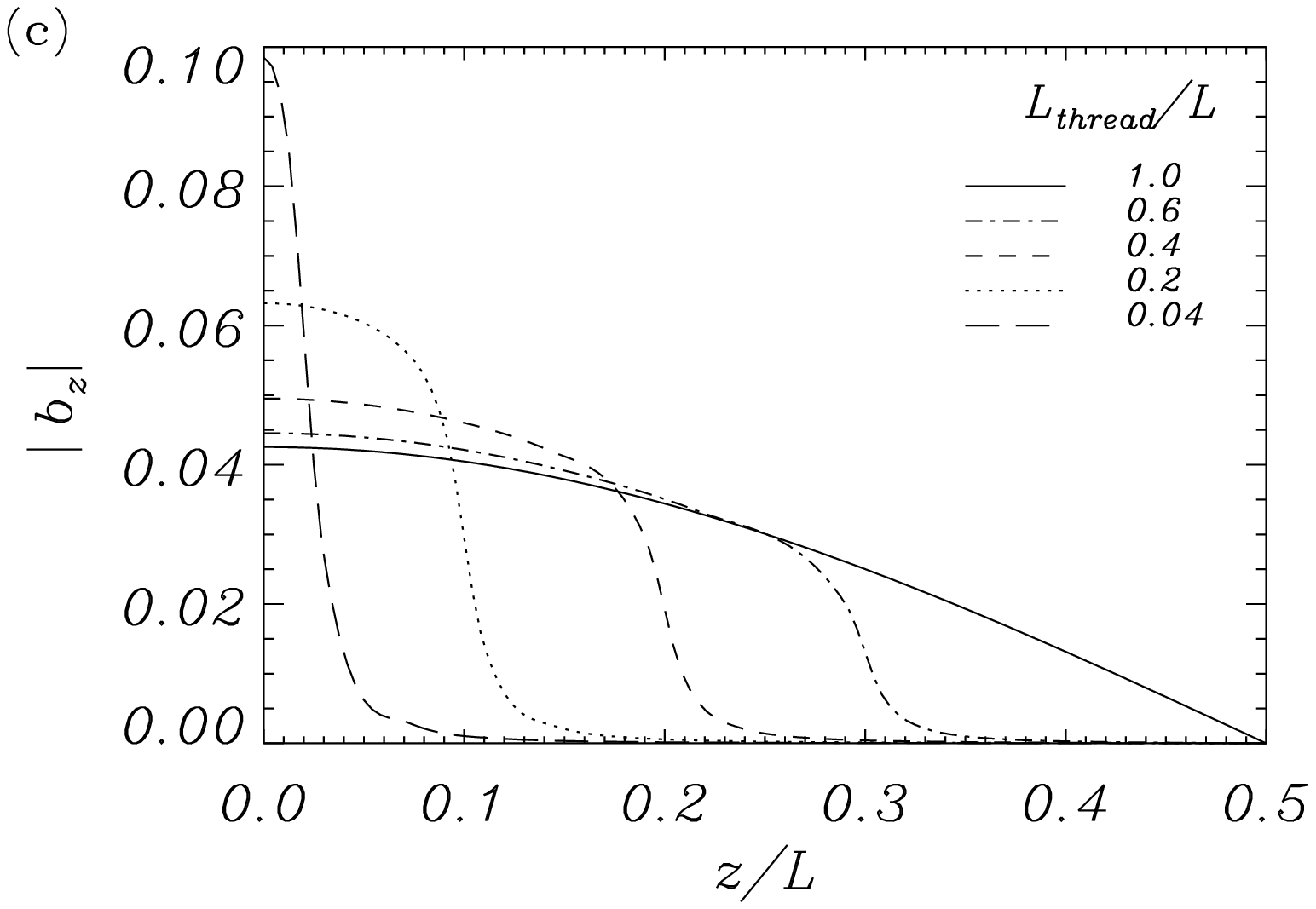}\\
 \includegraphics[width=5.8cm,height=4.55cm]{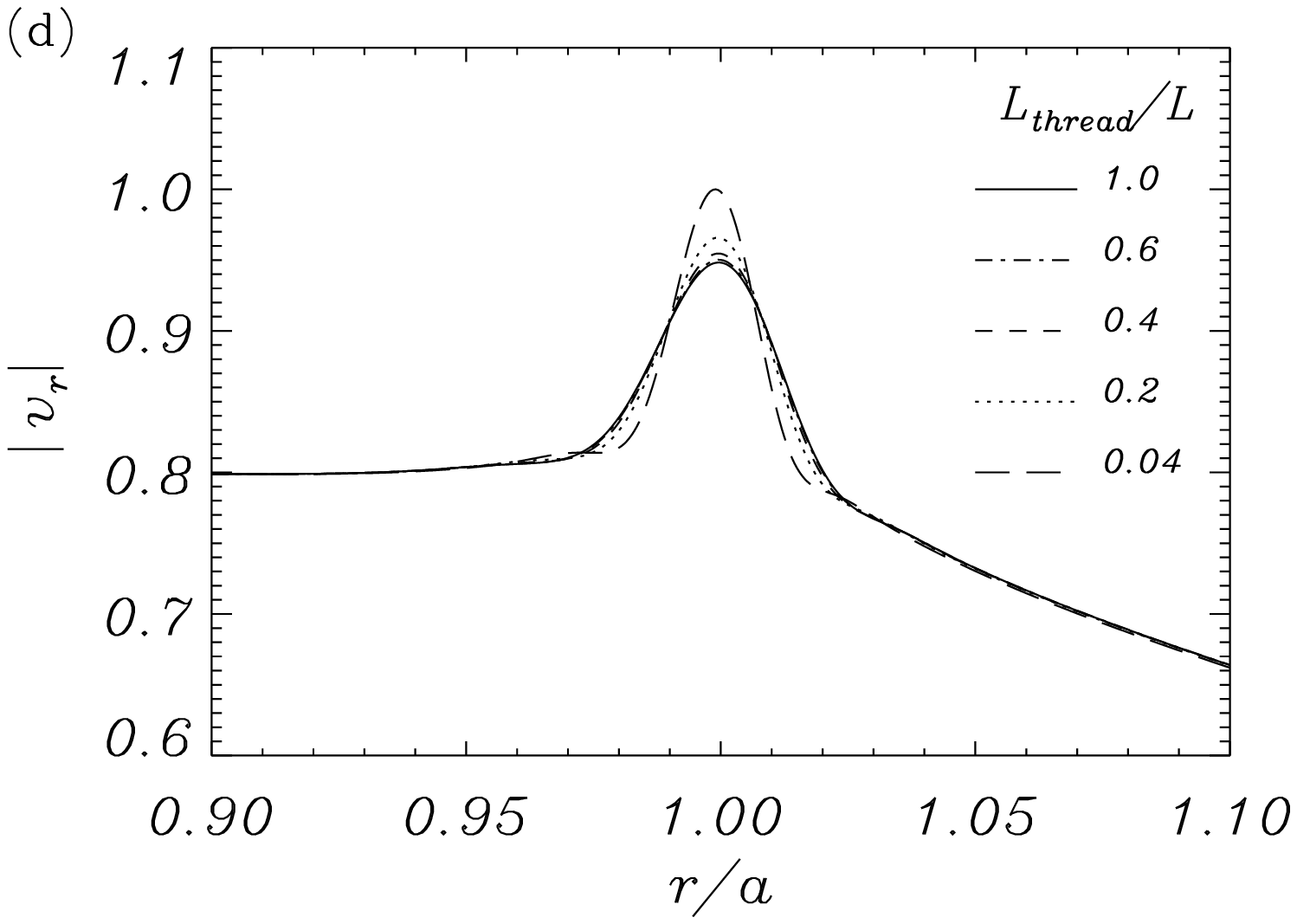}
 \includegraphics[width=5.8cm,height=4.55cm]{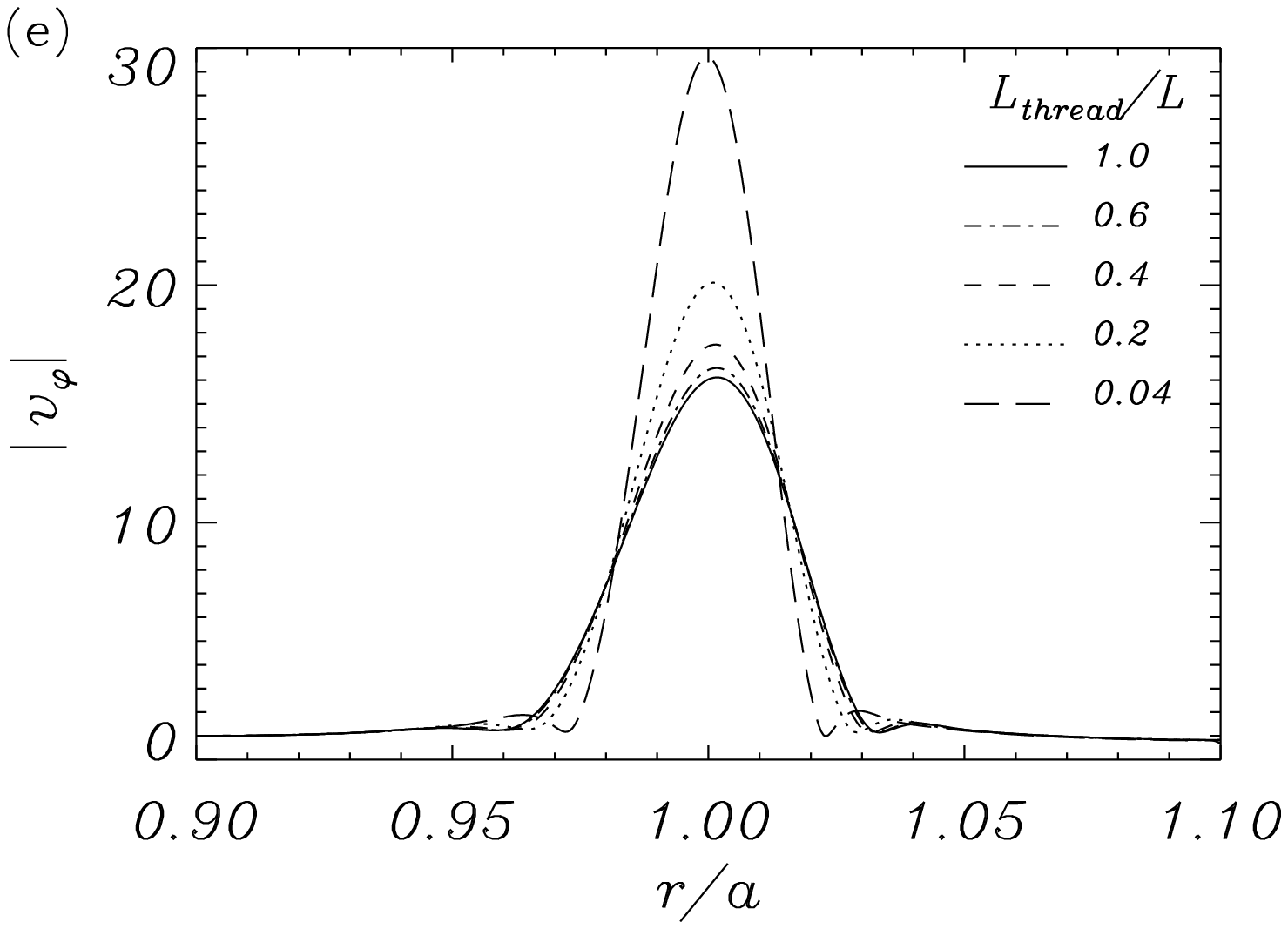}
 \includegraphics[width=5.8cm,height=4.55cm]{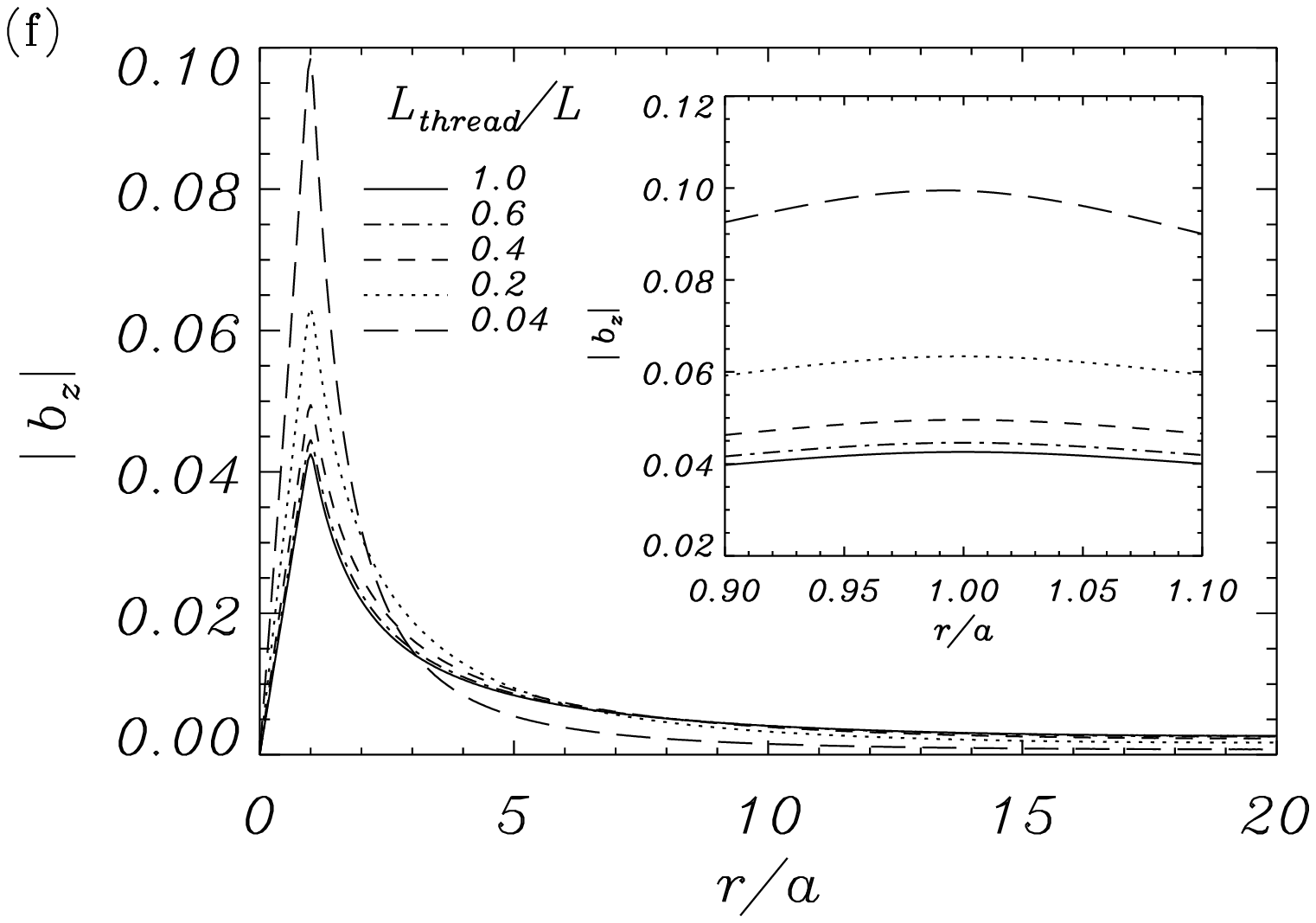}
 
\caption{Longitudinal and radial dependence of the eigenfunctions transverse velocity component, $v_r$, azimuthal velocity component, $v_\varphi$, and  compressive magnetic field component, $b_z$, in prominence threads with $\rho_{\rm f}/\rho_{\rm c}=200$, $l/a=0.2$, $a=1$, $\rho_{\rm ev}=\rho_{\rm c}$, and $R_{\rm m}=10^6$, for different values of the length of the thread. All computations have been performed in a two-dimensional grid with $N_r=401$ and $N_z=51$ points, with 250 grid-points in the resonant layer. \label{eigenlthread}}
\end{figure*}

First, we consider that $\rho_{\rm c} \leq \rho_{\rm ev} \leq \rho_{\rm f}$, hence the parameter is allowed to vary in between the coronal and prominence densities. Fig.~\ref{figrhoev} displays the obtained results. For $\rho_{\rm ev}=200\rho_{\rm c}=\rho_{\rm f}$, the tube is fully 
filled with cool and dense plasma. Then, as we decrease $\rho_{\rm ev}$, periods and damping times have a marked linear decrease. When 
$40\%$ of the tube is filled with cool plasma, they decrease by about a 15\%. When a 10\% filled tube is considered a decrease of up to a 50\%  
is obtained.  In physical terms, the period decrease when the density in the evacuated part of the tube is gradually decreased can be understood if we think about the different inertia of the system, with or without dense plasma at those locations along the tube. For explaining the different damping time-scales, energy arguments that combine the energy of the mode and the energy flux into the resonance have to be considered, see Sect.~\ref{energyanal}. The decrease in period and damping time is very similar, but not exactly the same. For instance, Fig.~\ref{figrhoev}c shows that 
the damping ratio is slightly dependent on the density in the evacuated part of the tube, for all the considered values in the range  
$\rho_{\rm c} \leq \rho_{\rm ev} \leq \rho_{\rm f}$.

Next, we have considered the possibility of the density in the evacuated part of the tube being 
lower than the coronal density. For instance, \citet{diaz02} considered a value of $\rho_{\rm ev}=0.6\rho_{\rm c}$. The computations shown in 
Fig.~\ref{figrhoev} are extended to lower values of $\rho_{\rm ev}$. The obtained results are displayed in the inset plots of Fig.~\ref{figrhoev} 
and show that the density in the evacuated part of the tube is irrelevant in relation to the period of the oscillations and the damping by resonant 
absorption in the considered range of values with $0.1\rho_{\rm c}\leq\rho_{\rm ev}\leq\rho_{\rm c}$.

These results show that  the density in the evacuated part of the tube is also a relevant parameter for prominence thread seismology,  because of its  
effect on periods and damping times and the difficulty in being measured by direct means. When considered in combination with effects due to the length of the thread, out computations enable us to perform a more accurate prominence seismology, applicable to a large number of threads.

\subsection{Spatial distribution of eigenfunctions}\label{spatial}

Changes in the longitudinal density structuring have a direct impact on the oscillatory period, which is easy to understand, but also on the damping time by resonant absorption.  Besides obtaining the parametric behaviour of kink mode periods and damping times as a function of the longitudinal density structuring, we aim to explain the obtained results. As a first step, we have analysed the spatial structure of eigenfunctions. The relevant perturbed quantities are the radial and azimuthal  velocity components, $v_r$ and $v_{\varphi}$, and the compressive component of the perturbed magnetic field, $b_z$, directly related to the magnetic pressure perturbation, $P_T=Bb_z/\mu$. 

We first consider the influence of the length of the thread on the profiles of the eigenfunctions in the radial and longitudinal directions. Results are 
given in terms of the modulus of the complex eigenfunctions. Fig.~\ref{eigenlthread} shows one-dimensional cuts along the longitudinal 
and radial directions of the eigenfunctions for different values of the length of the thread. The longitudinal profiles are shown at the axis ($r=0$) for $v_r$, and at
the mean radius of the tube ($r=a$) for $v_{\varphi}$ and $b_z$. In the longitudinal direction all three eigenfunctions 
display a trigonometric dependence with $z$, when $L_{\rm thread}=L$, i.e., the case that mimics the one-dimensional thread.  When the length of the 
thread is decreased several interesting effects occur. First, both perturbed velocity components display a slightly improved confinement. The maximum 
values of the velocity perturbations still occur in the dense part of the tube, but the drop-off rate changes, becoming almost linear outside the thread so 
that they satisfy the boundary condition at the foot-point of the tube. When eigenfunctions are normalised to the value of $|v_r|$ at the apex of the tube, 
it is seen that the decrease of the length of the thread produces larger amplitudes of the azimuthal velocity component (see Fig.~\ref{eigenlthread}b), 
related to the Alfv\'enic character of the mode. This amplitude almost doubles its value when going from $L_{\rm thread}=L$ to  $L_{\rm thread}=0.04L$. 
The $z$-component of the magnetic field perturbation gives an indication of the compressibility of the normal mode. Our results indicate that when the 
length of the thread is decreased, the longitudinal profile of $b_z$ (Fig.~\ref{eigenlthread}c) becomes strictly confined to the dense part of the tube, where it reaches its 
maximum value. As the thread gets shorter, the maximum value of $b_z$ increases, hence the compressibility becomes larger at the apex of the tube
for shorter threads oscillations, though the kink mode remains being an almost incompressible wave mode. Kink modes in fully filled magnetic flux tubes 
are almost incompressible. Longitudinal density structuring by the inclusion of a dense central part increases (decreases) kink mode  compressibility in the dense (evacuated) parts of the tube, in comparison to the one-dimensional flux tube kink modes.

\begin{figure*}

 \includegraphics[width=5.8cm,height=4.55cm]{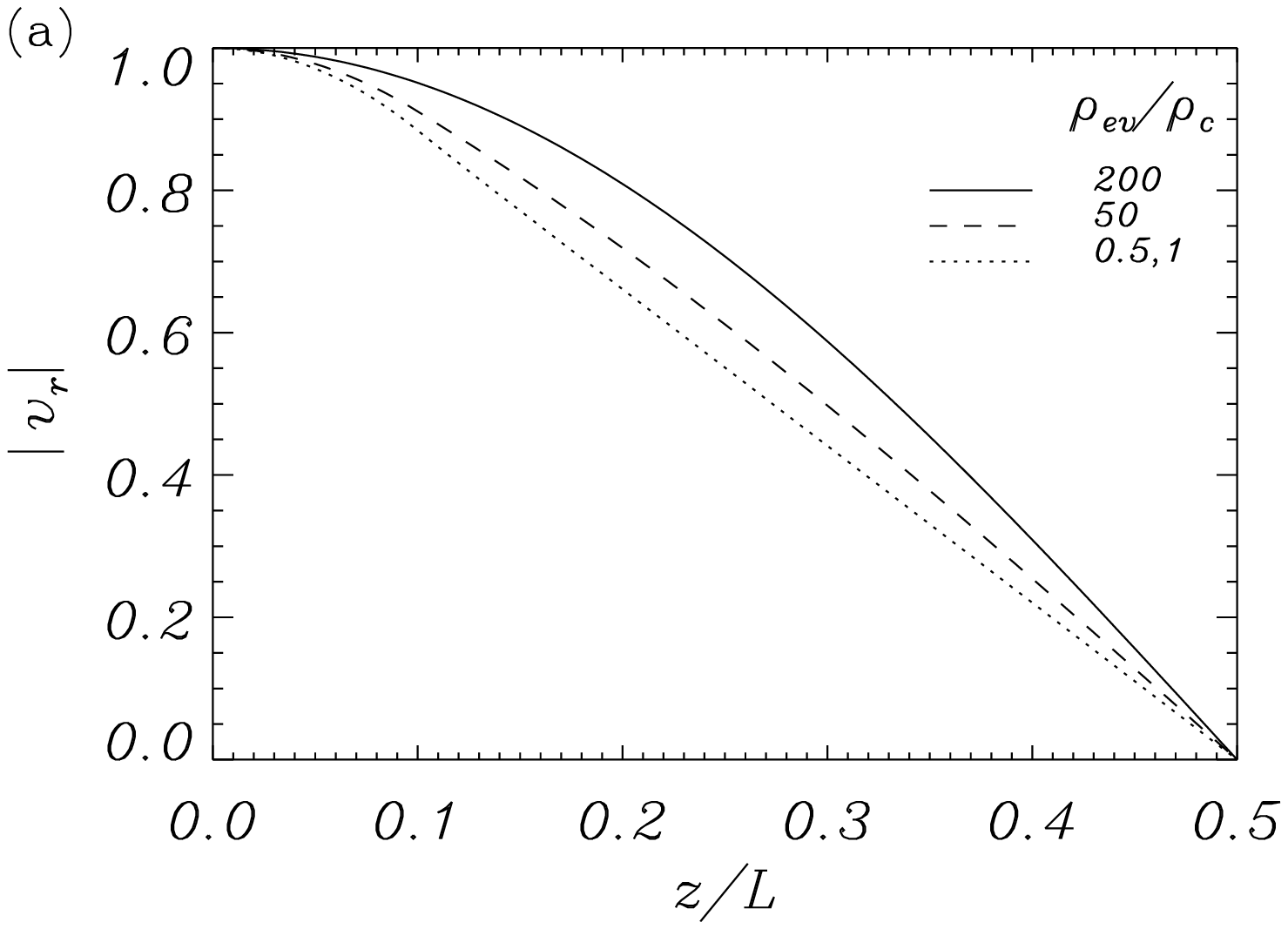}
 \includegraphics[width=5.8cm,height=4.55cm]{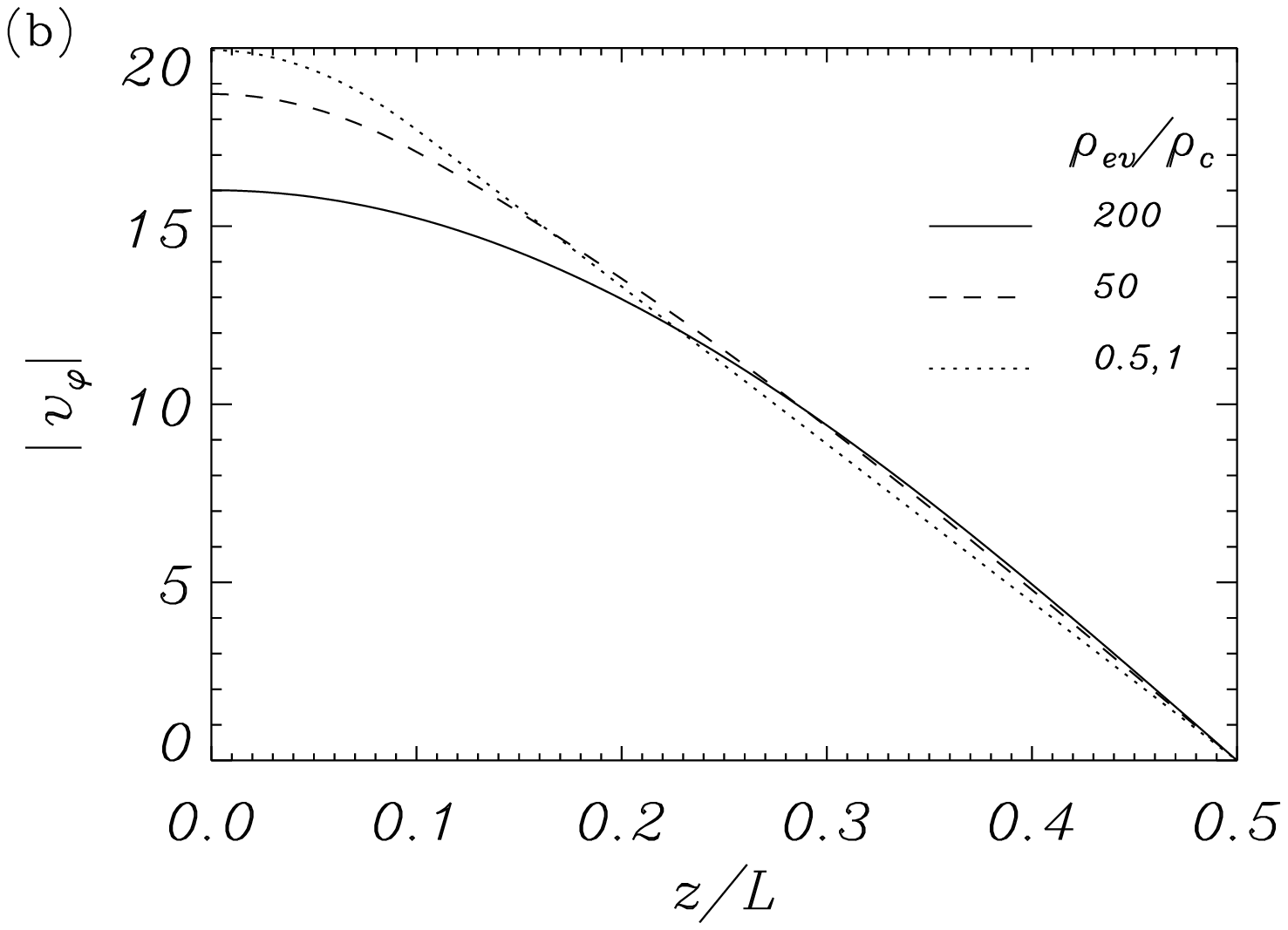}
 \includegraphics[width=5.8cm,height=4.55cm]{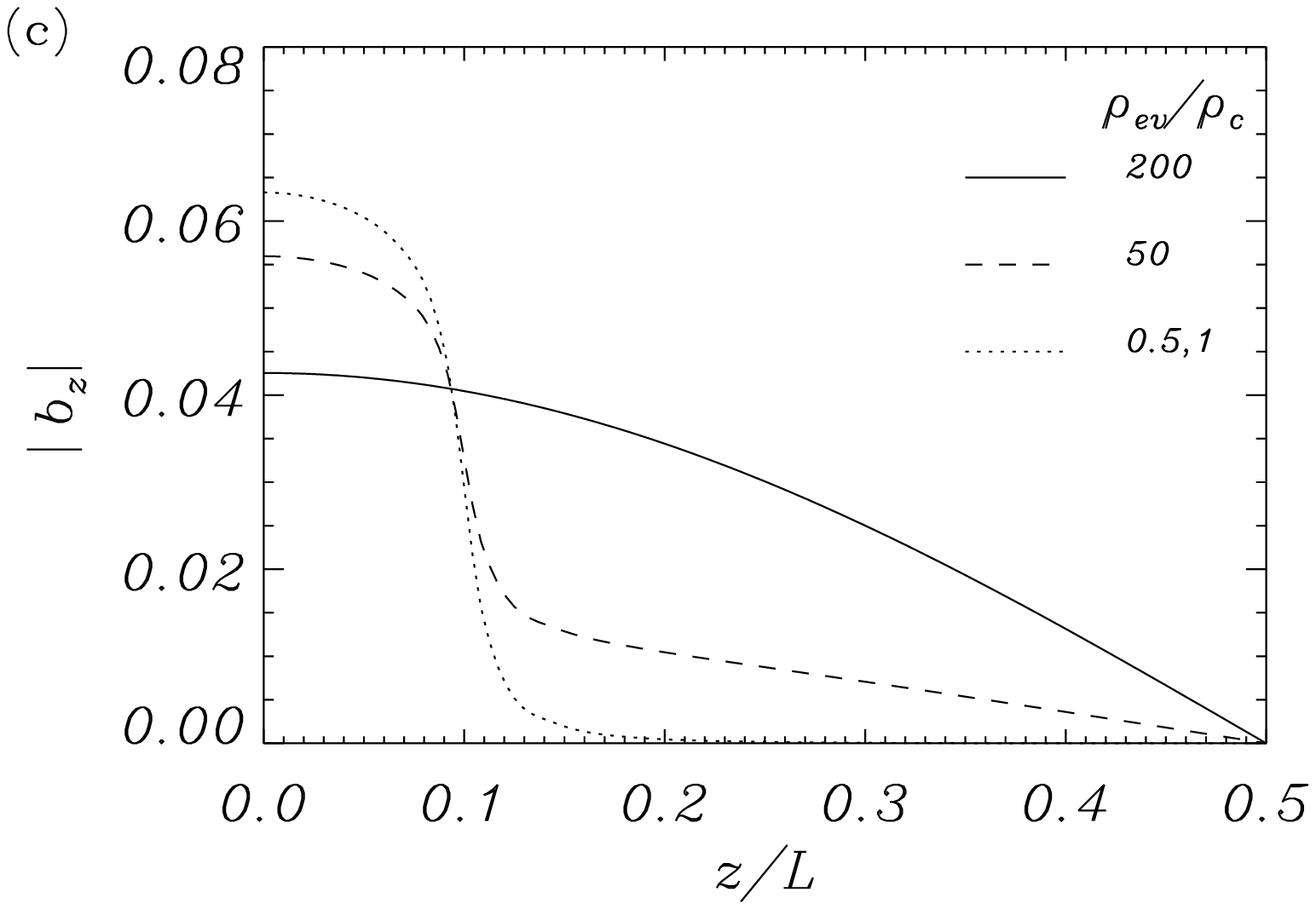}\\
 \includegraphics[width=5.8cm,height=4.55cm]{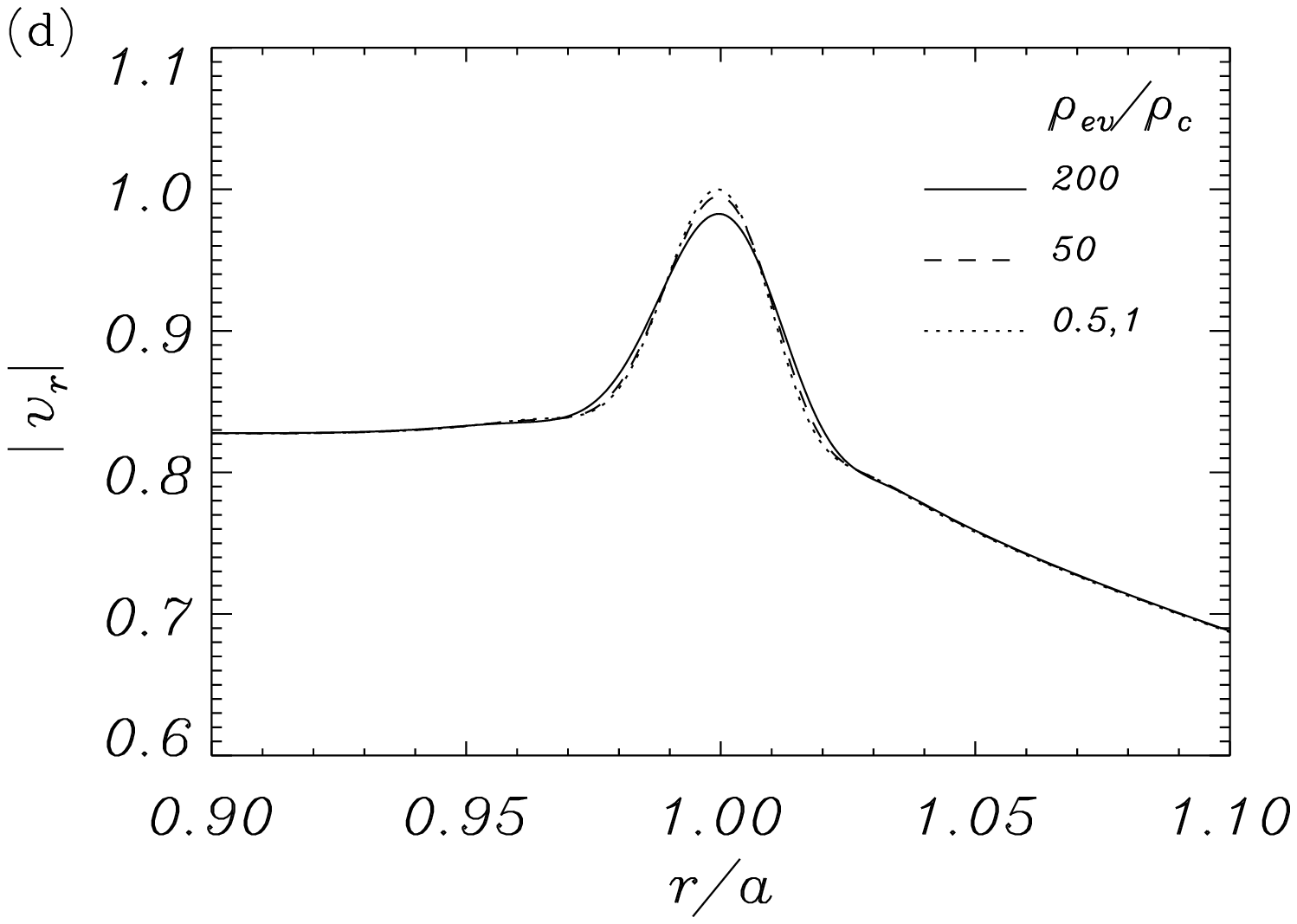}
 \includegraphics[width=5.8cm,height=4.55cm]{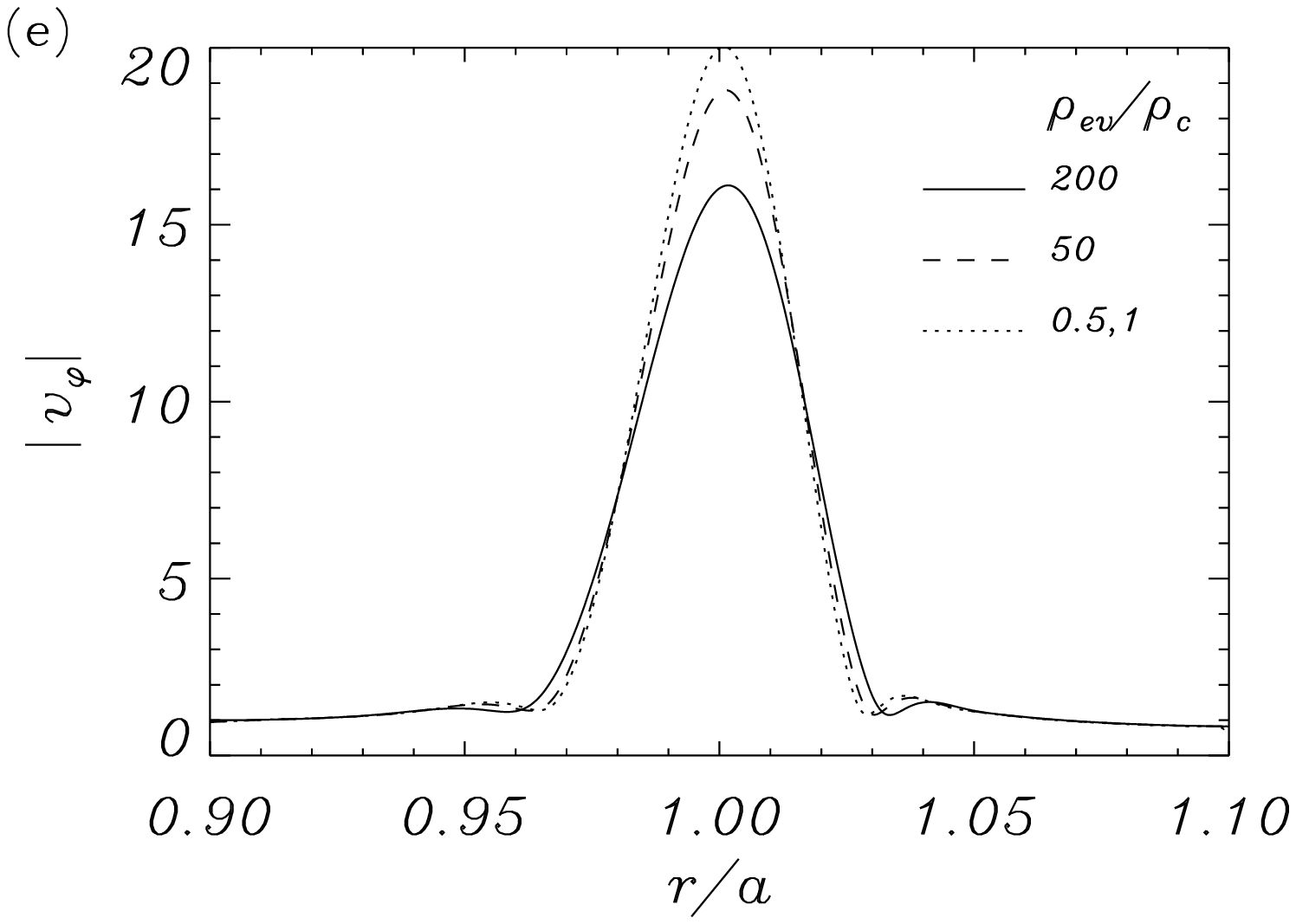}
 \includegraphics[width=5.8cm,height=4.55cm]{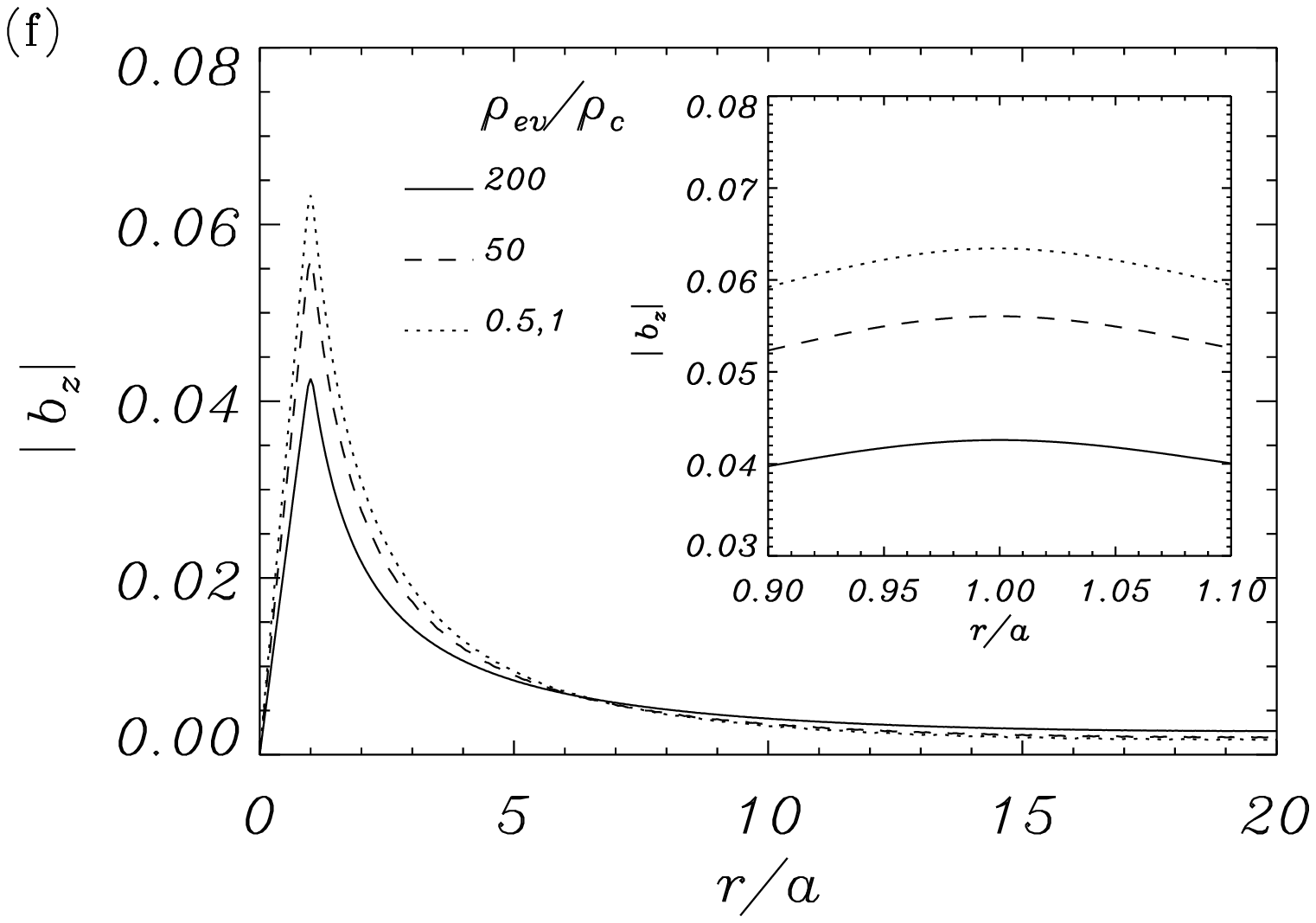}

\caption{Longitudinal and radial dependence of the eigenfunctions transverse velocity component, $v_r$, azimuthal velocity component, $v_\varphi$, and compressive magnetic field component, $b_z$ in prominence threads with $\rho_{\rm f}/\rho_{\rm c}=200$, $l/a=0.2$, $a=1$, $L_{\rm thread}=0.2L$, and $R_{\rm m}=10^6$ for different values of the density in the evacuated part of the tube. All computations have been performed in a two-dimensional grid with $N_r=401$ and $N_z=51$ points, with 250 grid-points in the resonant layer. \label{eigenrhoev}}
\end{figure*}

A close look at the spatial structure of eigenfunctions in the radial direction (Figs.~\ref{eigenlthread}d-f) provides us with a qualitative explanation on 
why the change of the length of the thread affects the damping times computed in the previous subsection (and also those presented by \citealt{soler102dthread}). First, the change in the period produced by the change in the length of the thread affects the resonant position in the radial direction and hence the damping time. However, the slope of the density profile in the transitional layer is very large, because of the high density contrast ratio typical of filament threads and this effect is not very important. Second, the eigenfunctions in the resonant layer are affected by the value of the length of the thread. Both effects are clearly seen in 
Figs.~\ref{eigenlthread}d-f, where the radial dependence of $v_r, v_\varphi$, and $b_z$ is plotted for different thread lengths. 
We first note that the perturbed velocity components at the resonance have a larger amplitude the shorter the length of the thread. We interpret this as an 
indication of an improved efficiency of the resonant damping mechanism. This is particularly clear if we look at the azimuthal velocity profile in the 
resonant layer (Fig.~\ref{eigenlthread}e). Now, not only the increase of the amplitude for shorter lengths is evident, but also the transverse length 
scale is seen to decrease when shorter threads are considered, hence thinner resonance widths are obtained. This detail of the resonance also shows 
a slight shift towards the left hand side which is due to the change of the resonant position for different lengths of the threads (hence different oscillatory 
periods). Although the resonances are not exactly located at $r_{\rm A}=a$, our numerical results indicate that this approximation, used 
by \citet{soler102dthread}, is fully justified. Finally, the radial profile of the compressive component of the magnetic field perturbation (Fig.~\ref{eigenlthread}f) shows the increase in its amplitude at the resonant position mentioned above, when considering shorter threads.  This is another indication of the improved efficiency of resonant damping.  Notice that although our study is limited to linear MHD waves, in a realistic situation perturbations inside the resonant layer become nonlinear, as shown in studies by e.g., \cite{Terradasnonlinear,Clack09,Ballai11}.

We have next examined the spatial distribution of the relevant perturbed quantities as a function of the density in the evacuated part 
of the tube. Fig.~\ref{eigenrhoev} shows the obtained profiles in the longitudinal and radial directions for a fixed value for the length of the thread. As before, the longitudinal profiles are shown at the axis ($r=0$) for $v_r$, and at the mean radius of the tube ($r=a$) for $v_{\varphi}$ and $b_z$. Our results indicate that the density in the evacuated part of the tube has also a direct impact on the radial and longitudinal profiles of eigenfunctions. For the longitudinal profiles, as we decrease the value of $\rho_{\rm ev}$ from $\rho_{\rm f}$, we have a slightly improved confinement of the velocity perturbations and a strict confinement of the longitudinal component of the perturbed magnetic field to the dense part  of the tube (see Figs.~\ref{eigenrhoev}a-c). The amplitude of $b_z$ at the apex of the tube increases, while in the evacuated part of the tube, and for a fixed length of the thread, compressibility decreases as $\rho_{\rm ev}$ is decreased (see Fig.~\ref{eigenrhoev}c).
Taking a look at the eigenfunctions in the radial direction (Figs.~\ref{eigenrhoev}d-f), the amplitude of both perturbed velocity components increases when the density in the evacuated part of the tube is decreased. The increase in the resonance peak and the shortening of the transverse spatial scale are mainly evident for the azimuthal velocity component, which gives its resonantly damped and Alfv\'enic character to the mode.  We find  an increase in the compressibility of the normal mode (Fig.~\ref{eigenrhoev}f) in the dense prominence plasma region as the density in the evacuated part of the tube is decreased. 

Overall, the decrease of the density in the evacuated part of the tube, starting from a fully filled tube, produces similar qualitative effects on the radial and longitudinal profiles for the eigenfunctions as the ones produced by the shortening of the length of the thread. In the parameter range studied in this work those effects are quantitatively more important in the case of the changes of the length of the thread. The properties of the spatial structure of eigenfunctions give us a qualitative explanation on why changes in the longitudinal density structuring in prominence threads have a significant effect on the damping time of kink oscillations in two-dimensional thread models.

\subsection{Energy analysis}\label{energyanal}

Our analysis of the spatial structure of eigenfunctions indicates that the decrease of both  the length of the thread and the density in the 
evacuated part of the tube  produce a strengthening of the resonance  absorption process that becomes apparent through the appearance of 
more pronounced resonant profiles and thinner resonance widths in the perturbed velocity components at the resonance. This would explain the marked 
decrease of the damping times found in Section~\ref{freqs} when varying these two parameters. This result shows that  resonant damping may strongly depend on the details of the longitudinal density structuring in rather general two-dimensional density models, such 
as the ones used here to model prominence threads.  

In order to obtain a quantitative explanation an energy analysis is carried out. Our analysis involves the 
energy of the kink mode and the energy flux into the resonance. The energy of the mode can directly be computed from the numerical eigenfunctions as

\begin{equation}
E=\frac{1}{2}\left(\rho v^2+b^2\right).\label{energykink}
\end{equation} 

\noindent
This expression has to be integrated over the entire ($r,z$)- plane except for the resonance layer, easily recognisable from the eigenfunctions displayed in Figs.~\ref{eigenlthread} and \ref{eigenrhoev}, as this region would include a contribution from the Alfv\'en waves. In one-dimensional equilibrium models, the energy flux into the resonance is proportional to the magnetic pressure perturbation squared  \citep{ATG00,AG01},  a result that was used by \cite{ATOB07a} to analyse the influence of the internal structuring of coronal loops on the damping by comparing the efficiency of the process at internal and external layers.  In two-dimensional equilibrium states, the energy flux absorbed at a particular field
line is proportional to the overlap integral between $P_T(r_{\rm A},z)$, the profile of $P_T$  along the tube at the resonant position, and the resonant Alfv\'en eigenfunctions \citep{TW93,TirryJGR}.  This is why the longitudinal profiles of $b_z$  and $v_\varphi$, and their modification due to  changes in the longitudinal density distribution are so relevant in determining kink mode damping times.  A general mathematical expression, valid 
in the three-dimensional case, is given by \cite{WT94}. When adapted to our cylindrical equilibrium, this expression gives the energy flux at the resonance per unit $\varphi$ in
the form

\begin{figure}
\includegraphics[width=7.8cm,height=5.55cm]{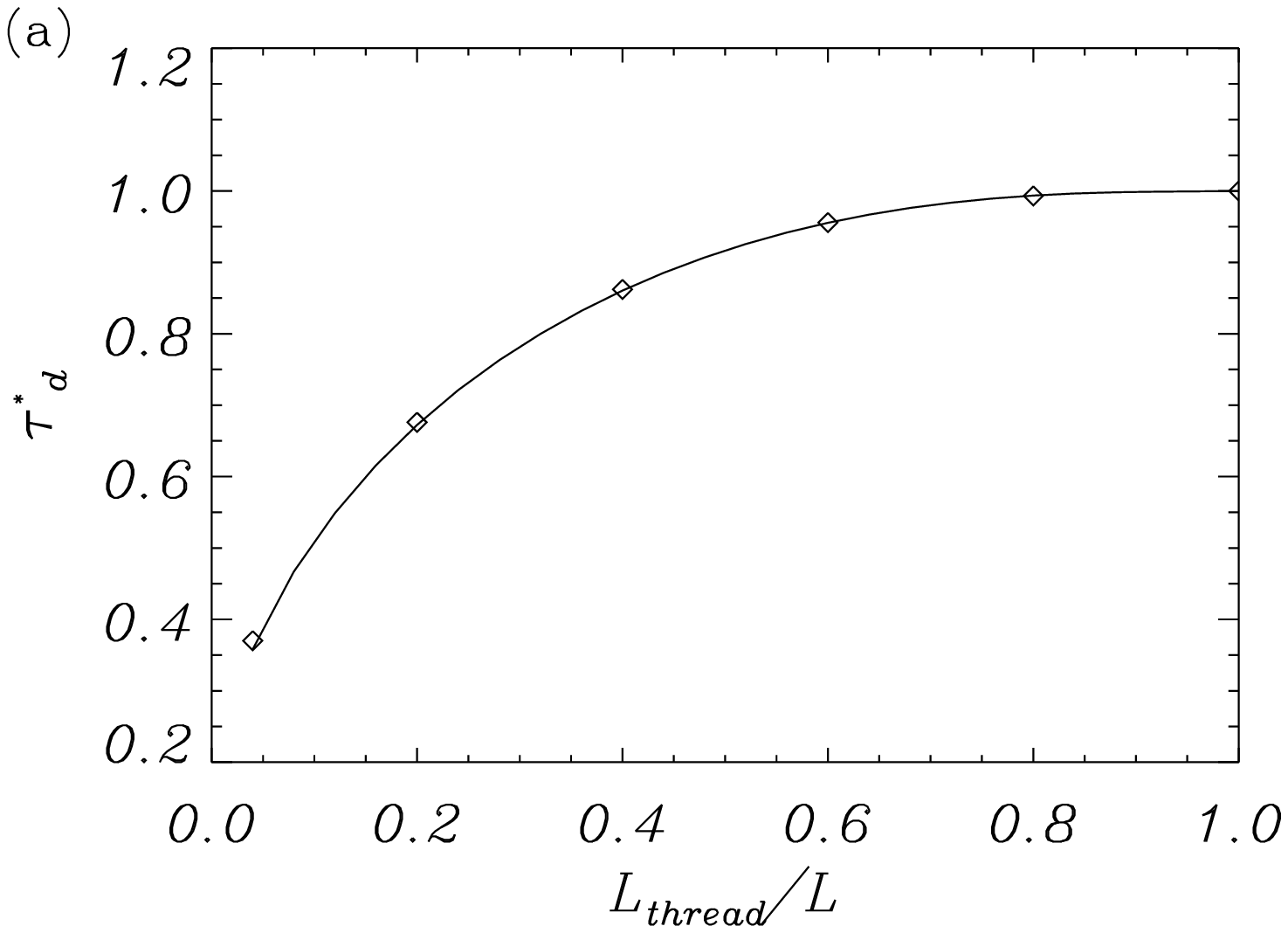}
 \includegraphics[width=7.8cm,height=5.55cm]{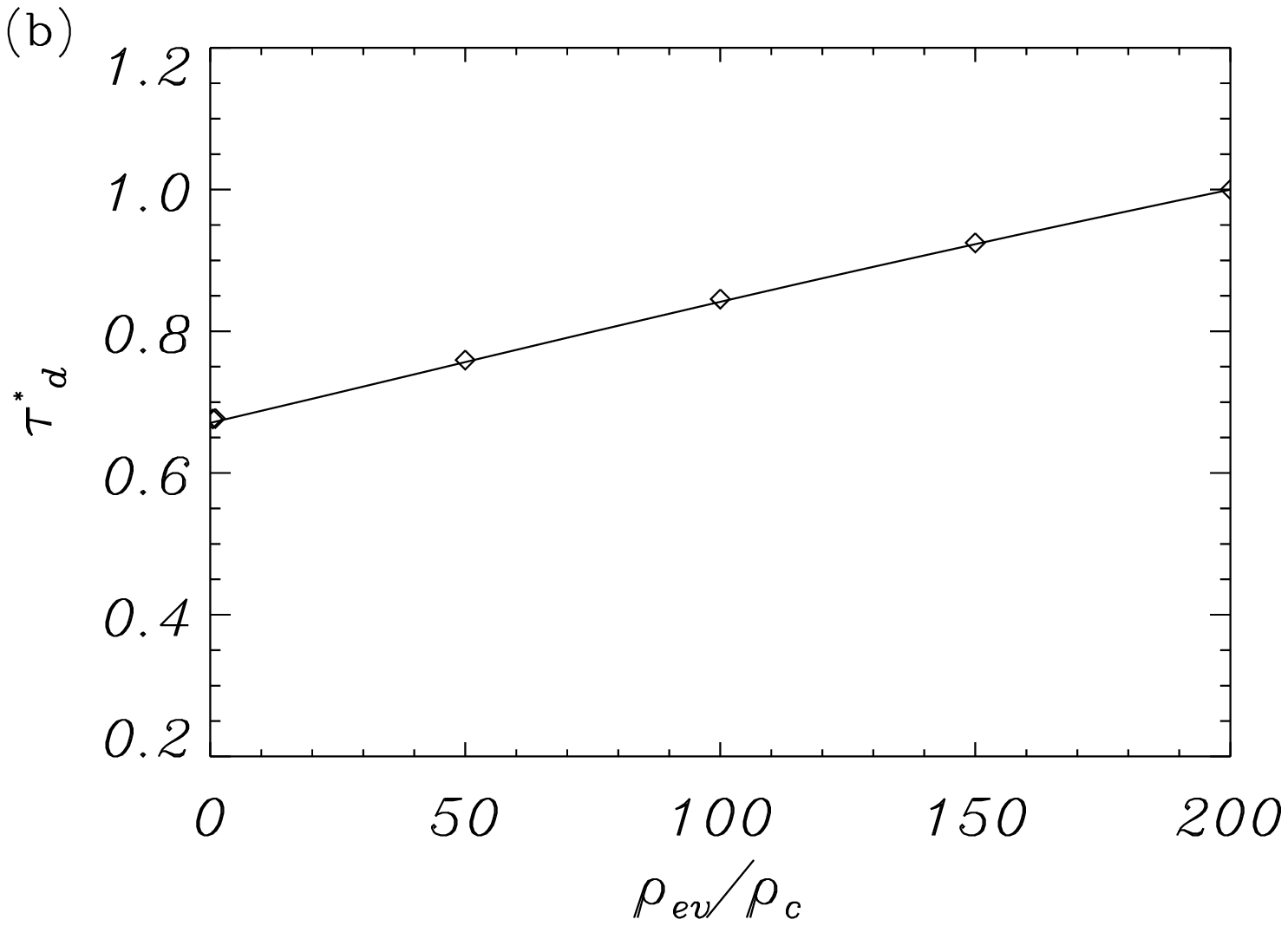}

\caption{(a) Damping time, normalised to the full tube damping time, as a function of the length of the thread, for $\rho_{\rm ev}=\rho_{\rm c}$. (b) Damping time, normalised to the full tube damping time, as a function of the density in the evacuated part of the tube, for $L_{\rm thread}=0.2L$. In both figures lines correspond to the numerically computed damping times and the symbols are the values obtained through the energy analysis described in Section~\ref{energyanal}. \label{dampingflux}}
\end{figure}

\begin{figure}
\includegraphics[width=7.8cm,height=5.55cm]{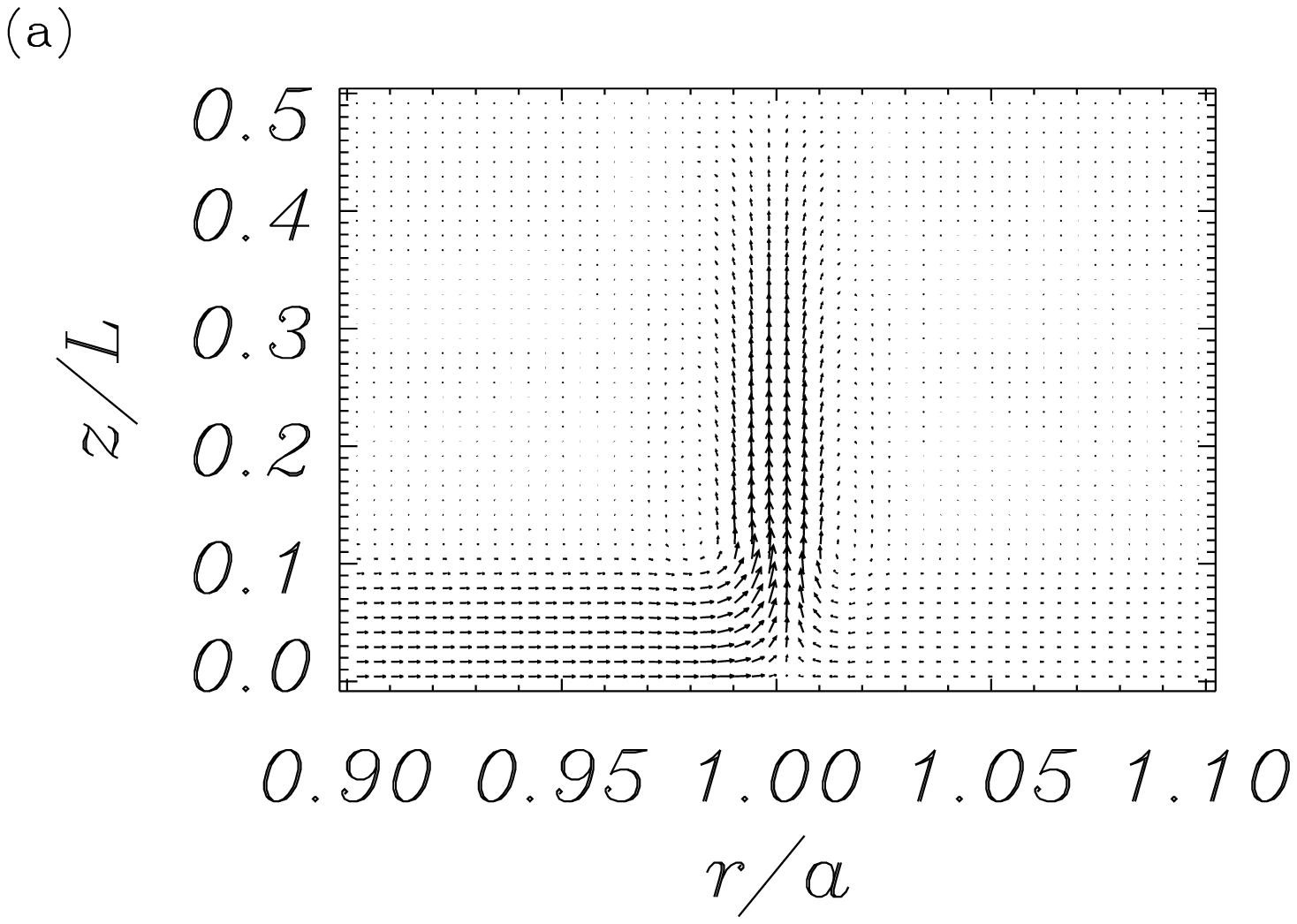}
 \includegraphics[width=7.8cm,height=5.55cm]{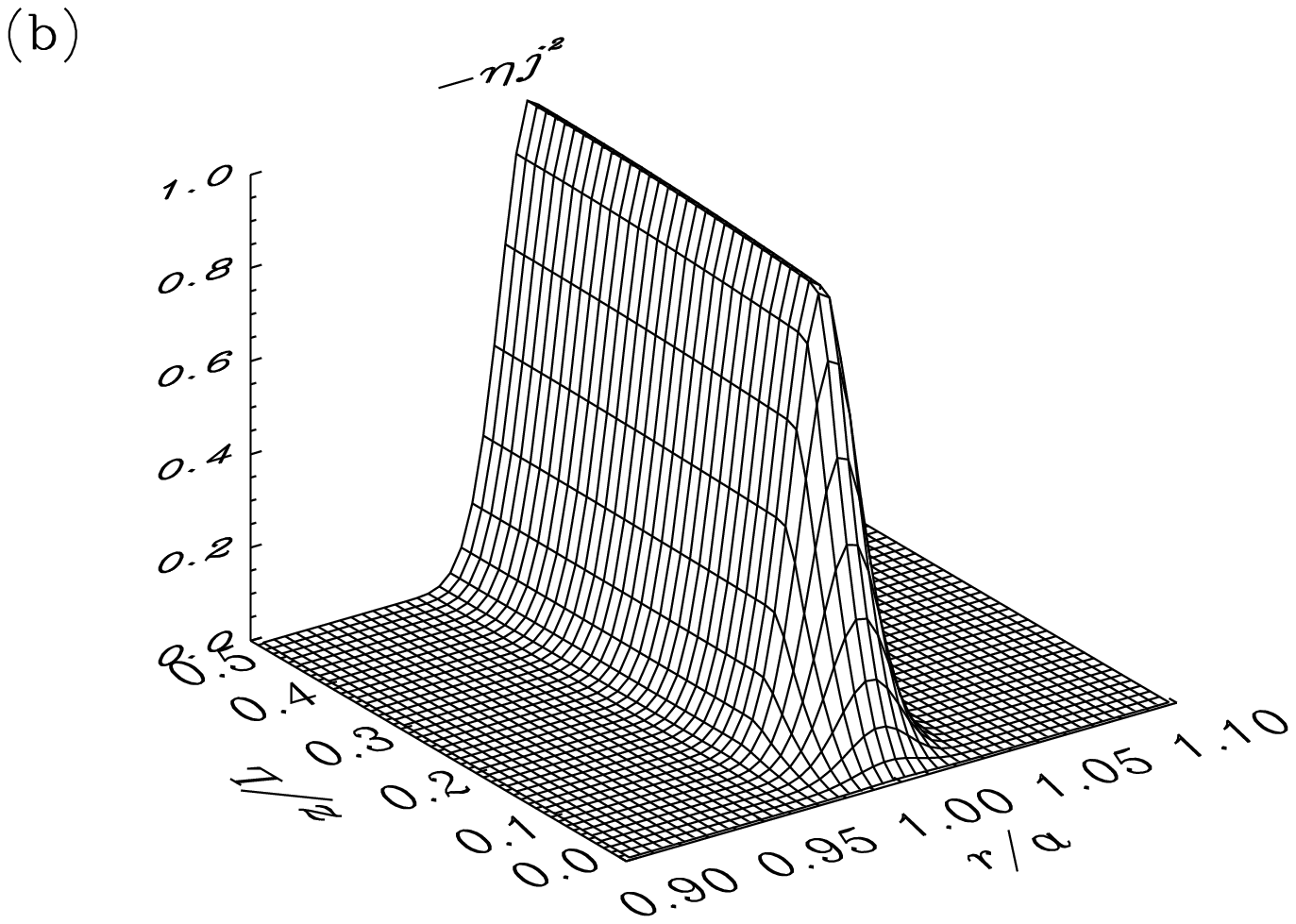}
\caption{(a) Spatial distribution in the ($r, z$)-plane of the time-averaged Poynting flux given by Equations~(\ref{sr}) and (\ref{sz}). (b) Ohmic heating distribution, in arbitrary units. These results correspond to the transverse oscillation of a prominence thread with $\rho_{\rm f}/\rho_{\rm c}=200$,  $\rho_{\rm ev}=\rho_{\rm c}$, $l/a=0.2$, $a=1$, and $L_{\rm thread}=0.2L$. The magnetic Reynolds number is $R_{\rm m}=10^6$.} 
\label{poyntingflux}
\end{figure}

\begin{eqnarray}\label{flux}
F&=&r_A\int \left[S_r(r^-_A)-S_r(r^+_A)\right] dz\\\nonumber
&=&\frac{\pi m^2}{4\mu_0^2}\times\frac{B^2}{r_A}\frac{(\int\phi \ b_z \ dz)^2}{\int\phi \ \rho \ \phi \  dz}\times\left(\frac{d\omega_A}{dr}\right)_{r_A}^{-1}.
\end{eqnarray} 

\noindent
In this expression, $S_r(r^-_A)$  and $S_r(r^+_A)$ represent the values of the radial component of the time-averaged Poynting vector to the left and to the right of the resonance position, that we approximate by $r_A=a$,  $\phi$ is the Alfv\'en eigenfunction along the tube at the resonant position, and $\omega_{\rm A}(r)$ the Alfv\'en continuum frequency.  
In order to evaluate the integrals as well as the slope of the Alfv\'en continuum at the resonant position in expression~(\ref{flux}), 
we have first solved the following equation for the Alfv\'en continuum modes

\begin{equation}
\frac{d^2\phi(r_A,z)}{dz^2}+\frac{\omega^2_A(r_A)}{v^2_A(r_A,z)}\phi(r_A,z)=0, 
\label{alfveneq}
\end{equation}

\noindent
with boundary conditions

\begin{displaymath}
\frac{d\phi}{dz}(z=0)=0,  \phi(z=L/2)=0.
\end{displaymath}
\noindent
For each value of $r_{\rm A}$, this equation is solved for different values for the relevant parameters $L_{\rm thread}$ and $\rho_{\rm ev}$, that in turn define different profiles for $v^2_A(r_A,z)$. The Alfv\'en continua  arise from repeating the procedure for different values of $r_A$ in the range $[a-l/2,a+l/2]$. Once this is done, the slope of the Alfv\'en continua at the resonant position and the Alfv\'en eigenfunctions that correspond to each thread model can be evaluated. 
By using the numerically computed profiles for $b_{\rm z}(r_{\rm A},z)$, in order to evaluate the overlap integral in Equation~(\ref{flux}), and multiplying by $2\pi$, the required total energy flux is obtained.  The ratio of the energy of the kink mode to the time-averaged energy flux into the resonance gives the required damping time, for every considered value of $L_{\rm thread}$ and $\rho_{\rm ev}$.

Figure~\ref{dampingflux} shows the obtained results. They have been normalised to the value of the damping time for the fully filled tube. We see an excellent agreement between the damping times computed through the energy analysis explained in detail above and the numerically computed ones for both cases in which we change the length of the thread and the density in the evacuated part of the tube. The results obtained in Section~\ref{freqs} can therefore be explained in terms of the energetics of the modes and the resonant energy transfer. These results also show the accuracy and utility of the analytical expression derived by \cite{WT94} for the time-averaged energy flux in terms of eigenfunctions.  Furthermore, we have just demonstrated that damping time estimates can be obtained, using energy arguments, without the need to solve the full non-uniform problem, but instead by solving the much more simpler piece-wise uniform problem in order to obtain the real part of the frequency and the longitudinal profiles for the eigenfunctions for kink modes in combination with the non-uniform computations for the Alfv\'en continuum modes, using this information in combination with Equations~(\ref{energykink}) and (\ref{flux}). This is possible because  the real part of the frequency and the longitudinal profiles of the eigenfunctions are only slightly affected by resonant coupling, and because the energy in Equation~(\ref{energykink}) has to be integrated excluding the resonant layer. Of course, the full numerical solution provides us with more accurate results, but as 
Fig.~\ref{dampingflux} illustrates the agreement between both methods is excellent.

A convenient way to further understand the energy flux at the resonance and the dynamics of resonantly damped kink modes is to compute 
the spatial distribution of the time-averaged Poynting flux in our two-dimensional domain by making use of the general expression

\begin{equation}
<{\bf S}>=\frac{1}{2}\mathrm{Re}({\bf E}\times{\bf b^*}),
\end{equation}

\noindent
with ${\bf E}=-({\bf v}\times{\bf B})+\eta{\bf j}$ the perturbed electric field,  ${\bf b^*}$=$(b^*_r, b^*_\varphi, b^*_z)$ the complex conjugate of the perturbed magnetic field, and ${\bf j}=(\nabla\times{\bf b})/\mu$ the current. In contrast to the energy analysis above, we now make use of the full numerical computations. In terms of the wave fields analysed in Section~\ref{spatial}, the two components of the time-averaged Poynting vector in the ($r$, $z$)- plane can be cast as

\begin{eqnarray}
<S_{\mathrm r}> &=&\frac{1}{2}\frac{B}{\mu}\mathrm{Re}(v_r b^*_z)+\frac{\eta}{2\mu^2} \mathrm{Re}\left[ b^*_z \left(\frac{\partial b_r}{\partial z}-\frac{\partial b_z}{\partial r}\right)\right.\nonumber\\
&-& \left. b^*_\varphi \left(\frac{\partial b_\varphi}{\partial r}+\frac{b_\varphi}{r}-\frac{im}{r} b_r\right)\right],\label{sr}\\
<S_{\mathrm z}> &=&-\frac{1}{2}\frac{B}{\mu}\mathrm{Re}(v_r b^*_r+v_\varphi b^*_\varphi)+\frac{\eta}{2\mu^2} \mathrm{Re}\left[b^*_\varphi \left(\frac{im}{r}b_z-\frac{\partial b_\varphi}{\partial z}\right) \right.\nonumber\\
&-& \left. b^*_r \left(\frac{\partial b_r}{\partial z}-\frac{\partial b_z}{\partial r}\right)\right].\label{sz}
\end{eqnarray}

\noindent
By making use of the divergence-free condition given by Equation~(\ref{divb}) to compute $b_\varphi$, we have produced an example of the two-dimensional distribution of the time-averaged Poynting flux around the resonant layer, for a partially filled thread.

The arrow plot in Fig.~\ref{poyntingflux}a shows that energy is fed into the resonant layer by concentrating it over the dense thread section. 
The amount of energy flux in the radial direction is determined by the jump in $<S_{\rm r}>$  \citep[see][for a one-dimensional example in ideal MHD]{stenuit99}. 
The resistive layer is on a scale where both the ideal and resistive terms of the perturbed electric field can be of similar magnitude, but each of them could be dominant over different regions of the domain. For instance, the jump in $<S_r>$ is determined by the ideal term in Equation~(\ref{sr}) which is  dominant in the thread region and zero in the evacuated part of the tube. Hence the energy inflow towards the resonance in the radial direction comes from the interior of the tube and is determined by the spatial profiles of the transverse velocity component and the compressive magnetic field perturbation. Note that the amplitude of the kink mode is smaller outside the tube compared to inside and also that the evacuated part of the tube is much less dense than the thread.

Once in the layer,  the energy flow is diverted along the field lines in a manner determined by $<S_{\rm z}>$, with a small radial contribution that  is due to the resistive terms in Equation~(\ref{sr}).  Such as corresponds to the Alfv\'enic character of the mode inside the resonant layer, the dominant term in Equation~(\ref{sz}) involves the azimuthal velocity and magnetic field perturbations, $v_{\varphi} b^*_{\varphi}$. This quantity decreases linearly along the field lines, producing the lessening in the parallel Poynting flux towards the foot-point. In our example, the radial non-uniform layer is restricted to the dense part of the flux tube, since $\rho_{\rm ev}=\rho_{\rm c}$. However, energy concentrated in the resonant layer of the thread because of resonant wave damping can flow along the field lines and, eventually, supply heating in the evacuated region, where field aligned currents are dominant.  Although the energy inflow into the resonance, given by $<S_r>$, and its subsequent divergence along the field lines, given by $<S_z>$, are mostly determined by ideal terms in Equations~(\ref{sr}) and (\ref{sz}), this does not mean that resistivity is not important. For instance, the amount of heating, in the form of Ohmic dissipation, will be determined by those currents, resistivity, and their spatial distribution. For this particular case, heating is distributed in a constant manner in the evacuated part of the tube, even if  there is no resonant layer in that region (see Fig.~\ref{poyntingflux}b).

\section{Summary and Conclusions}\label{sect4}

Quiescent filament fine structures are only partially filled with cold and dense absorbing material. The length of the threads can in principle be 
measured in events showing transverse oscillations, provided the lifetime of threads is sufficiently large compared to the oscillatory 
period. The length of the supporting magnetic flux tubes are however much larger, and cannot be observed. Density measurements, both 
in the thread as in the evacuated part of the supporting magnetic tube,  are also challenging from the observational point of view. It is  therefore 
important to quantify the variations on wave properties due to changes in these equilibrium parameters if we aim to perform an accurate prominence seismology. It is essential to have computations of periods and damping times for a wide range of thread models to include regimes in which 
the applicability of simple analytical models could be of limited extent. For this reason we have computed the oscillatory properties of resonantly damped transverse kink oscillations in rather general  two-dimensional fully non-uniform prominence thread models. This allows for a broad range of prominence threads with very different physical conditions to be modelled and their oscillatory properties characterised.

The length of the thread and the density in the evacuated part of the tube define their longitudinal density structuring. We find that the length of the 
thread strongly influences the period and damping time of transverse kink oscillations, while the damping ratio is rather insensitive to this parameter. 
These results confirm the validity of the analytical approximations made by \citet{soler102dthread}. In addition, our modelling has allowed us to 
identify a new physical parameter with seismological implications, the density in the evacuated part of the thread.  This quantity also influences periods 
and damping times, and to a lesser extent damping ratios, and must be taken into account in the inversion of physical parameters in the context of 
prominence seismology.

Currently available  inversion schemes for one-dimensional coronal loops and prominence threads \citep{Arregui07,GABW08,AB10,soler102dthread} make use of observed periods and damping ratios. The first, influence the inferred values for the Alfv\'en speed, while the second determine the transverse density structuring.
Based on our results,  we can conclude that ignorance on the length of the thread, the length of the supporting magnetic flux tube, and the density in the evacuated part of the tube will have a significant impact on the inferred values for the  Alfv\'en speed (hence magnetic field strength) in the thread,  
depending on whether we use those one-dimensional inversion schemes or the results from two-dimensional models here obtained. 
On the contrary, because of the smaller sensitivity of the damping ratio to changes in the longitudinal density structuring, seismological estimates of the transverse density structuring will be less affected by our ignorance about  the longitudinal density structuring of prominence threads.

 Our study provides additional insight to the physics of resonantly damped kink modes in two-dimensional equilibrium states, by extending previous
 applications \citep[e.g,][]{Andries05,Arregui05} to more complex non-separable density distributions.  It also provides an example of the methods  and uses of combining the information from the spatial distribution of eigenfunctions with that obtained from energy arguments. In particular,  our energy analysis has allowed us to explain the decrease in damping times for shorter thread lengths found by \cite{soler102dthread}.  The length of the thread influences the energy of the kink mode, and hence its oscillatory period, but also affects the damping by resonant absorption, through the energy flux into the resonance. In an analogous way, the value of the density in the evacuated part of the tube also determines periods and damping times, since both the energy of the kink mode and the energy flux into the resonance vary. This means that changes in the equilibrium configuration in a non-resonant direction produce variations in the damping properties of kink modes, a result that was qualitatively explained by a detailed examination of the radial and longitudinal profiles of the eigenfunctions. Both the shortening of the length of the thread and the decrease of the density in the evacuated part of the tube produce more marked resonances, with the amplitude of the velocity perturbations at the resonance and the compressibility of the mode in the thread being larger. Inside the resonant layers shorter transverse spatial scales for the Alfv\'enic velocity component are obtained. In combination with the analysis of the energy of the kink modes and the energy flux into the resonances a quantitative explanation was obtained for both the damping properties obtained in our study and those in \citet{soler102dthread}.

The damping of kink oscillations in two-dimensional fully non-uniform equilibrium configurations can be computed by using energy arguments together with the solution of simpler problems for kink mode and Alfv\'en continuum modes. This aspect is worth to be considered in future studies of resonant absorption in 2D/3D models of solar atmospheric magnetic structures involving changes of equilibrium parameters that affect the density structuring in a non-resonant direction. The use of energy arguments would allow to have a first indication about how important a given parameter that modifies the equilibrium in a non-resonant direction is, while avoiding to solve the full problem until we are interested in the details. 

The two-dimensional distribution of the time-averaged Poynting flux shows how energy is fed into the resonance and subsequently flows along the field lines by  properties of wave fields associated to ideal processes. This is the reason why resonant damping is a mechanism for wave energy transfer in which time-scales are independent of resistivity, in the limit of high magnetic Reynolds numbers. Magnetic diffusion plays its role once energy is concentrated at small spatial scales, by providing heating at locations that, as in our example  for a partially filled thread, are distributed in regions where no resonant layer is present. This result  offers additional insights to the dynamics of resonantly damped kink modes and the heating of prominence plasmas by wave transformation processes. It must be considered in detail and  extended to density models relevant to other solar atmospheric structures.

\begin{acknowledgements}
IA, RS, and JLB acknowledge the funding provided under the project AYA2006-07637 by Spanish MICINN and FEDER Funds and the discussions within the ISSI Team on Solar Prominence Formation and Equilibrium: New Data, New Models.  RS acknowledges support from a Marie Curie Intra-European Fellowship within the European Commission 7th Framework Program (PIEF-GA-2010-274716). We are grateful to Ram\'on Oliver and Jaume Terradas for a number of useful comments and suggestions.
\end{acknowledgements}


\end{document}